\documentclass[aps,prm,twocolumn,showpacs,psfig,superscriptaddress,longbibliography]{revtex4-1}
\usepackage{textcomp}
\usepackage{times}
\usepackage{graphicx}
\usepackage{float}
\usepackage{latexsym,amsmath,amssymb,bm,euscript}
\usepackage{color}
\usepackage{subfigure}
\usepackage{epstopdf}
\usepackage[colorlinks=true,linkcolor=blue,citecolor=blue]{hyperref}
\usepackage{hyperref}
\usepackage{soul}
\usepackage[normalem]{ulem}
\usepackage{mathrsfs}
\usepackage{amsmath}
\usepackage{lettrine}
\usepackage{xspace}
\usepackage{textcomp}

\DeclareGraphicsExtensions{.pdf, .png, .jpg, .eps}

\begin{document}

\title{Molecular intercalation in the van der Waals antiferromagnets FePS$_3$ and NiPS$_3$}

\author{Cong Li}
\thanks{These authors contributed equally to this study.}
\affiliation{Department of Physics and Beijing Key Laboratory of
Opto-electronic Functional Materials $\&$ Micro-nano Devices, Renmin
University of China, Beijing, 100872, China}

\author{Ze Hu}
\thanks{These authors contributed equally to this study.}
\affiliation{Department of Physics and Beijing Key Laboratory of  Opto-electronic Functional Materials $\&$ Micro-nano Devices, Renmin
University of China, Beijing, 100872, China}

\author{Xiaofei Hou}
\thanks{These authors contributed equally to this study.}
\affiliation{School of Physical Science and Technology, ShanghaiTech University, Shanghai, 201210, China}

\author{Sheng Xu}
\affiliation{Department of Physics and Beijing Key Laboratory of  Opto-electronic Functional Materials $\&$ Micro-nano Devices, Renmin
University of China, Beijing, 100872, China}

\author{Zhanlong Wu}
\affiliation{Department of Physics and Beijing Key Laboratory of  Opto-electronic Functional Materials $\&$ Micro-nano Devices, Renmin
University of China, Beijing, 100872, China}

\author{Kefan Du}
\affiliation{Department of Physics and Beijing Key Laboratory of  Opto-electronic Functional Materials $\&$ Micro-nano Devices, Renmin
University of China, Beijing, 100872, China}

\author{Shuo Li}
\affiliation{Department of Physics and Beijing Key Laboratory of  Opto-electronic Functional Materials $\&$ Micro-nano Devices, Renmin
University of China, Beijing, 100872, China}

\author{Xiaoyu Xu}
\affiliation{Department of Physics and Beijing Key Laboratory of  Opto-electronic Functional Materials $\&$ Micro-nano Devices, Renmin
University of China, Beijing, 100872, China}

\author{Ying Chen}
\affiliation{Department of Physics and Beijing Key Laboratory of  Opto-electronic Functional Materials $\&$ Micro-nano Devices, Renmin
University of China, Beijing, 100872, China}

\author{Zeyu Wang}
\affiliation{Department of Chemistry, Renmin University of China, Beijing, 100872, China}

\author{Tiancheng Mu}
\affiliation{Department of Chemistry, Renmin University of China, Beijing, 100872, China}

\author{Tian-Long Xia}
\email{tlxia@ruc.edu.cn}
\affiliation{Department of Physics and Beijing Key Laboratory of  Opto-electronic Functional Materials $\&$ Micro-nano Devices, Renmin
University of China, Beijing, 100872, China}
\affiliation{Key Laboratory of Quantum State Construction and Manipulation (Ministry of Education),
Renmin University of China, Beijing, 100872, China}

\author{Yanfeng Guo}
\email{guoyf@shanghaitech.edu.cn}
\affiliation{School of Physical Science and Technology, ShanghaiTech University, Shanghai, 201210, China}
\affiliation{ShanghaiTech Laboratory for Topological Physics, Shanghai, 201210, China}

\author{B. Normand}
\affiliation{Laboratory for Theoretical and Computational Physics, Paul Scherrer Institute, CH-5232 Villigen-PSI, Switzerland}
\affiliation{Institute of Physics, Ecole Polytechnique F\'ed\'erale de Lausanne (EPFL), CH-1015 Lausanne, Switzerland}

\author{Weiqiang Yu}
\email{wqyu\_phy@ruc.edu.cn}
\affiliation{Department of Physics and Beijing Key Laboratory of  Opto-electronic Functional Materials $\&$ Micro-nano Devices, Renmin
University of China, Beijing, 100872, China}
\affiliation{Key Laboratory of Quantum State Construction and Manipulation (Ministry of Education),
Renmin University of China, Beijing, 100872, China}

\author{Yi Cui}
\email{cuiyi@ruc.edu.cn}
\affiliation{Department of Physics and Beijing Key Laboratory of  Opto-electronic Functional Materials $\&$ Micro-nano Devices, Renmin
University of China, Beijing, 100872, China}
\affiliation{Key Laboratory of Quantum State Construction and Manipulation (Ministry of Education),
Renmin University of China, Beijing, 100872, China}


\begin{abstract}
We have performed electrochemical treatment of the van der Waals antiferromagnetic materials FePS$_3$
and NiPS$_3$ with the ionic liquid EMIM-BF$_4$, achieving significant molecular intercalation. 
Mass analysis of the intercalated compounds, EMIM$_x$-FePS$_3$ and EMIM$_x$-NiPS$_3$, indicated 
respective intercalation levels, $x$, of approximately 27\% and 37\%, and X-ray diffraction measurements 
demonstrated a massive (over 50\%) enhancement of the $c$-axis lattice parameters. 
To investigate the consequences of these changes for the magnetic properties, we performed magnetic 
susceptibility and $^{31}$P nuclear magnetic resonance (NMR) studies of both systems. 
For EMIM$_x$-FePS$_3$, intercalation reduces the magnetic ordering temperature from $T_N = 120$~K 
to 78~K, and we find a spin gap in the antiferromagnetic phase that drops from 45~K to 30~K. 
For EMIM$_x$-NiPS$_3$, the ordering temperature is almost unaffected (changing from 148~K to 145~K),
but a change towards nearly isotropic spin fluctuations suggests an alteration of the magnetic Hamiltonian. 
Such relatively modest changes, given that the huge extension of the $c$ axes is expected to cause a very 
strong suppression any interlayer interactions, point unequivocally to the conclusion that the magnetic 
properties of both parent compounds are determined solely by two-dimensional (2D), intralayer physics.   
The changes in transition temperatures and low-temperature spin dynamics in both compounds therefore 
indicate that intercalation also results in a significant modulation of the intralayer magnetic 
interactions, which we propose is due to charge doping and localization on the P sites. 
Our study offers chemical intercalation with ionic liquids as an effective method to control not 
only the interlayer but also the intralayer interactions in quasi-2D magnetic materials.
\end{abstract}

\maketitle

\section{\label{sintro}Introduction}

Low-dimensional magnetic systems have played a pivotal role not only in advancing our understanding of 
the quantum properties of materials but also in the realization and exploration of new concepts in 
many-body physics. Still, with the exception of certain geometrically discretized or strong-frustration 
scenarios, which lead to quantum disordered magnetic states including quantum spin liquids, most 
three-dimensional (3D) magnetic systems exhibit long-ranged order. Conversely, as emphasized by 
Bethe in his seminal work, low dimensions amplify quantum fluctuations and destabilize the conventional 
order parameters~\cite{Bethe_EPJA_1931}. In the case of layered compounds, increasing the interlayer 
spacing by the insertion of large molecules is a very literal means of modulating the dimensionality, 
and hence the properties, from potentially 3D towards the 2D limit~\cite{Joy_JACS_1992,Joy_ChemMater_1993}.
The discovery of 2D magnetic materials suitable for this type of control would hold significant promise
for applications in nanoelectronics and spintronics \cite{srfp_2016}.

For this reason, magnetic van der Waals materials have attracted extensive attention in recent years 
\cite{burch_nature_2018}, although efforts at dimensional manipulation have so far been limited largely 
to approaching the monolayer limit by exfoliation. In this context, the Ising-type van der Waals magnets 
Cr$_2$Ge$_2$Te$_6$ and CrI$_3$ exhibit the emergence of intrinsic ferromagnetism with a high transition 
temperature in few- or monolayer films~\cite{Gong_Nature_2017,Huang_Nature_2017,Tian_2D_2016}; in particular, 
CrI$_3$ shows a systematic evolution of the ordering temperature and even the type of magnetic order with 
the number of layers~\cite{Huang_Nature_2017}. MnSe$_2$ remains ferromagnetic (FM) at room temperature 
in its monolayer form~\cite{Hara_2018_nanoLett}, and most surprisingly room-temperature FM order 
emerges in monolayer VSe$_2$ despite the bulk material being paramagnetic~\cite{Boni_NatNon_2018}.

The transition-metal trisulfide $M$PS$_3$ ($M$~=~Mn, Fe, Co, Ni) is a class of antiferromagnetic (AFM) 
van der Waals materials that has also been studied extensively~\cite{Flem_JPCS_1982,Ouvrard_MRB_1985,
Joy_PRB_1992,Chittari_PRB_2016,Chris_ACSNano_2016,Wang_AdvML_2018}. As Fig.~\ref{structure}(a) shows, 
the $M$PS$_3$ compounds are isostructural, with the magnetic ions ($M^{2+}$) forming a honeycomb lattice.
All exhibit the characteristics of a Mott insulator, displaying high resistivity at room temperature 
and a band gap well in excess of 1~eV (1.5~eV for FePS$_3$ and 1.6~eV for NiPS$_3$). This gap varies 
widely on exfoliation, offering an optoelectronic response over a broad frequency range for device 
applications~\cite{Chris_ACSNano_2016}. FePS$_3$ exhibits a phase transition under pressure from 
insulating to metallic~\cite{Haines_PRL_2018} and NiPS$_3$ in its AFM phase exhibits coherent excitonic 
states \cite{kang_nature_2020}. 

Differences in the trigonal distortion of the $M$S$_6$ octahedra and in spin-orbit coupling mean that 
the anisotropy of intralayer magnetic interactions differs in the $M$PS$_3$ materials. MnPS$_3$ appears 
to have Heisenberg spin interactions, NiPS$_3$ shows a weak and CoPS$_3$ a stronger easy-plane (XY) 
anisotropy, and FePS$_3$ has a strong Ising anisotropy~\cite{Yasuo_1983,Joy_PRB_1992,Pramana_1994, Chatterjee_PRB_1995,Hicks_PRB_2007,Lee_NL_2016,Wildes_JPCM_2017}. Upon cooling, they all order in an 
AFM pattern, with respective transition temperatures, $T_N$, of 78~K, 118~K, 122~K, and 155~K for 
MnPS$_3$, FePS$_3$, CoPS$_3$ and NiPS$_3$~\cite{Flem_JPCS_1982,BREC_SSI_1986,Wang_AdvML_2018}.

The weak van der Waals interlayer coupling leads to a cleavage energy close to that of graphite, and 
thus the $M$PS$_3$ materials are easy to exfoliate. Multiple efforts to thin $M$PS$_3$ samples have 
shown them to remain structurally stable down to a single layer~\cite{Chris_ACSNano_2016,Kuo_SciRep_2016}.
Recent Raman scattering studies of monolayer FePS$_3$ have reported that $T_N$ either drops from 117~K to
104~K~\cite{Wang_2DMaterial_2016} or remains unchanged~\cite{Lee_NL_2016}, which raises the possibility 
of substrate effects on this 2D Ising magnet~\cite{Walker_PRB_2020}. Raman investigation of monolayer 
NiPS$_3$ indicates a suppression of long-range order relative to the bulk material, which was interpreted 
in the framework of the Berezinskii-Kosterlitz-Thouless (BKT) phase transition~\cite{Hu_PRB_2023}.

An alternative approach to dimensionality reduction is the incorporation of organic cations as spacers 
in the bulk materials, such that the altered layer separation enables control over the interlayer 
interactions~\cite{Zhou_SB_2020,Zhou_NP_2022}. This is complementary to electrochemical treatments with 
ionic liquids whose primary effect is to induce protonation, or other electron doping effects, when the 
sample is placed on the cathode side~\cite{cui_ScienceBulletin_2018,Cui_CPL_2019}. It was reported in the 
early literature that pyridine (C$_5$H$_5$N) can be intercalated into MnPS$_3$, leading to a transition
from an AFM to a weakly FM ground state~\cite{Joy_JACS_1992}, and that the intercalation of alkylamines 
(C$_n$H$_{2n+1}$NH$_2$) into FePS$_3$ causes a strong reduction of $T_N$~\cite{Joy_ChemMater_1993}. Much 
more recently, the intercalation of tetraheptylammonium-bromide (C$_{28}$H$_{60}$BrN) into NiPS$_3$ was 
found to cause a transition from AFM to ferrimagnetic (FIM) order, followed by another transition to AFM 
order, in effects ascribed to the electron doping of the layers \cite{Mi_AdvFM_2022}.

In this work, we report the successful interlayer intercalation of EMIM$^+$ into FePS$_3$ and NiPS$_3$, 
using the ionic liquid C$_6$H$_{11}$N$_2$BF$_4$ (EMIM-BF$_4$). By measuring changes in the mass and 
interlayer spacing, we estimate the intercalation levels to be approximately 0.27 EMIM$^+$/f.u.~in 
FePS$_3$ and 0.37 EMIM$^+$/f.u.~in NiPS$_3$, and that both procedures dilate the $c$-axis lattice 
parameter by over 50\%. We investigated the magnetic properties of EMIM$_x$-FePS$_3$ and EMIM$_x$-NiPS$_3$, 
and compared them with FePS$_3$ and NiPS$_3$, by magnetic susceptibility and $^{31}$P nuclear magnetic 
resonance (NMR) measurements.We found that intercalation in FePS$_3$ causes changes in the Curie-Weiss 
temperatures determined above $T_N$, a suppression of $T_N$ itself, and a reduced spin gap at the lowest 
temperatures. For intercalated NiPS$_3$, the magnetic transition temperature barely changes and the spin fluctuations 
become very isotropic, suggesting an evolution from weakly XY toward Heisenberg-type magnetism. Taken 
together with a dramatic suppression of any interlayer magnetic interactions, these data provide strong 
evidence for the systematic modulation of intralayer magnetic interactions by intercalation, for which we  
deduce that charge doping on the P site should be taken into account.

The structure of this article is as follows. In Sec.~\ref{stech} we describe our $M$PS$_3$ samples and 
intercalation procedure, and in Sec.~\ref{sstruc} report our structural characterization. Section 
\ref{sfeps3} reports our measurements of the magnetic properties of pure and intercalated FePS$_3$ and 
Sec.~\ref{snips3} our data for pure and intercalated NiPS$_3$. In Sec.~\ref{sdiscu} we discuss the 
consequences of our results for the understanding of magnetism in $M$PS$_3$ materials and provide 
a brief conclusion. 

\section{\label{stech} Materials and Methods}

\begin{figure}[t]
\includegraphics[width=8.6cm]{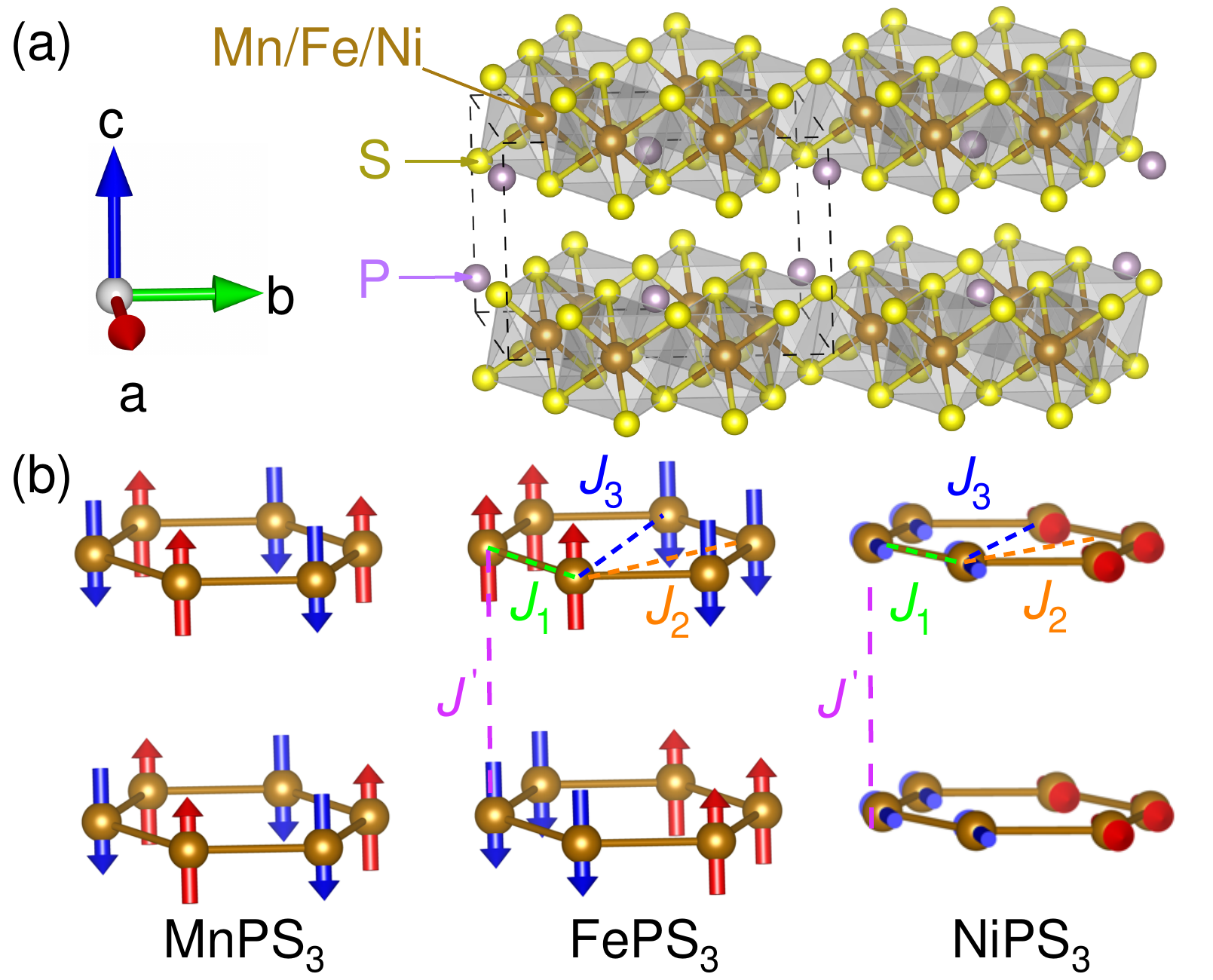}
\caption{\label{structure} \textbf{Lattice and magnetic structure of $M$PS$_3$.}
(a) Crystal structure. The figure illustrates two unit cells containing eight $M$PS$_3$ units.
(b) Magnetic structure below $T_N$. $J_1$, $J_2$, and $J_3$ label magnetic interactions between 
nearest-, next-nearest-, and next-next-nearest-neighbor sites, respectively; $J'$ labels the 
interlayer interaction, which we note corresponds to ABC stacking of the magnetic ions in the 
honeycomb layers. Blue and red arrows illustrate the relative orientations of local moments in 
the ordered phases.}
\end{figure}

Single crystals of FePS$_3$ and NiPS$_3$ were grown by the method of chemical vapor transport 
(CVT)~\cite{BREC_SSI_1986}. As Fig.~\ref{structure}(a) shows, the $M$PS$_3$ layer is composed 
of covalently bonded $M$S$_6$ octahedra and double-cone [P$_2$S$_6$]$^{4-}$ units in a 2:1 
ratio~\cite{brec_MRB_1985}. The $M$ ions are thought to be in a robustly divalent $M^{2+}$ 
state in all compounds, as shown for FePS$_3$ by X-ray photoelectron spectroscopy (XPS) 
\cite{Guo_JMCA_2019}. These ions form the  honeycomb lattice, with the P-P pairs located at 
the centers of the $M$ hexagons~\cite{BREC_SSI_1986,Joy_PRB_1992,Flem_JPCS_1982}. The S$^{2-}$ 
layers stack in an ABC configuration along the $c$ direction to form a monoclinic structure 
with space group C2/m~\cite{Ouvrard_MRB_1985}.

Both FePS$_3$ and NiPS$_3$ order magnetically with a ``zig-zag'' pattern, as shown in 
Fig.~\ref{structure}(b). The Fe$^{2+}$ ions in FePS$_3$ take their high-spin, $S = 2$ state 
and order with the moments normal to the $ab$~plane~\cite{Joy_PRB_1992}. These moments have FM 
alignment parallel to the $a$ axis and AFM alignment along the $b$ and $c$ axes, with propagation 
wave vector $\bm{k} = [0~1~\frac12]$~\cite{Yasuo_1983,Hicks_PRB_2007,Wildes_JPCM_2012,
Wildes_PRB_2016,Coak_PRX_2021,paul_2023_tuning}. For NiPS$_3$, the Ni$^{2+}$ ($S = 1$) moments 
are oriented primarily along the $a$ axis with a small component perpendicular to the $ab$ plane; 
the zig-zag chains also lie along the $a$ axis, with AFM alignment along $b$ but an FM repeat 
along the $c$ axis, resulting in $\bm{k} = [0~1~0]$~\cite{Yasuo_1983,Wildes_PRB_2015,Kim_NL_2021,
Wildes_PRB_2022}.

Figure~\ref{xrd}(a) shows the configuration that we adopt for electrochemical treatment in order 
to achieve interlayer intercalation. The ionic liquid EMIM-BF$_4$ was packed in a container with 
two platinum electrodes placed approximately 20~mm apart and subjected to a 3~V potential 
difference~\cite{Cui_CPL_2019,cui_ScienceBulletin_2018}. A single crystal of FePS$_3$ or NiPS$_3$ 
was attached to the cathode and covered with silver paint. The ionic liquid was heated to a 
temperature around 60$^\circ$C. After 24~hours the mass of the samples had changed significantly, 
but longer treatment times led to no further change, and so one could declare the intercalation 
process to be complete. 

X-ray diffraction (XRD) measurements were performed using Cu$_\alpha$ and Cu$_\beta$ radiation 
to determine the $c$-axis lattice parameters. The d.c.~magnetic susceptibility was measured 
in a Magnetic Property Measurement System (MPMS) with a field of 100~Oe and at temperatures down to 
1.8~K in field-cooled (FC) and zero-field-cooled (ZFC) conditions. $^{31}$P has nuclear spin $I = 1/2$ 
and a Zeeman factor of $^{31}\gamma = 17.235$~MHz/T. The $^{31}$P NMR measurements were performed on 
single crystals using a top tuning circuit and the spectra collected by spin-echo pulse sequences. 
For broad spectra, frequency sweeps were used to acquire the full spectrum. NMR Knight shifts were 
calculated from $K_n = (f/^{31}\gamma H - 1) {\times} 100\%$, where $f$ is the resonance frequency of 
the spectral peaks in the paramagnetic phase and the average frequency in the ordered phase. Spin-lattice 
relaxation rates, $1/^{31}T_1$, were measured by the standard magnetization inversion-recovery method. 
These constitute a sensitive probe of low-energy spin fluctuations, because $1/T_1 T = {\rm lim}_{\omega 
\to 0} \Sigma_q A^2_{\rm hf} (q) \, {\rm Im} \, \chi (q,\omega)/\omega$, where $\chi (q, \omega)$ is the 
dynamical susceptibility, $A_{\rm hf} (q)$ is the hyperfine coupling, and $\omega$ is the NMR 
measurement frequency, which for electronic spins lies in the zero-energy limit.

\section{\label{sstruc}Lattice structure and doping of EMIM$_{\lowercase{x}}$-F\lowercase{e}PS$_3$ and 
EMIM$_{\lowercase{x}}$-N\lowercase{i}PS$_3$  
\label{sec:struc}}

\begin{figure}[t]
\includegraphics[width=8.6cm]{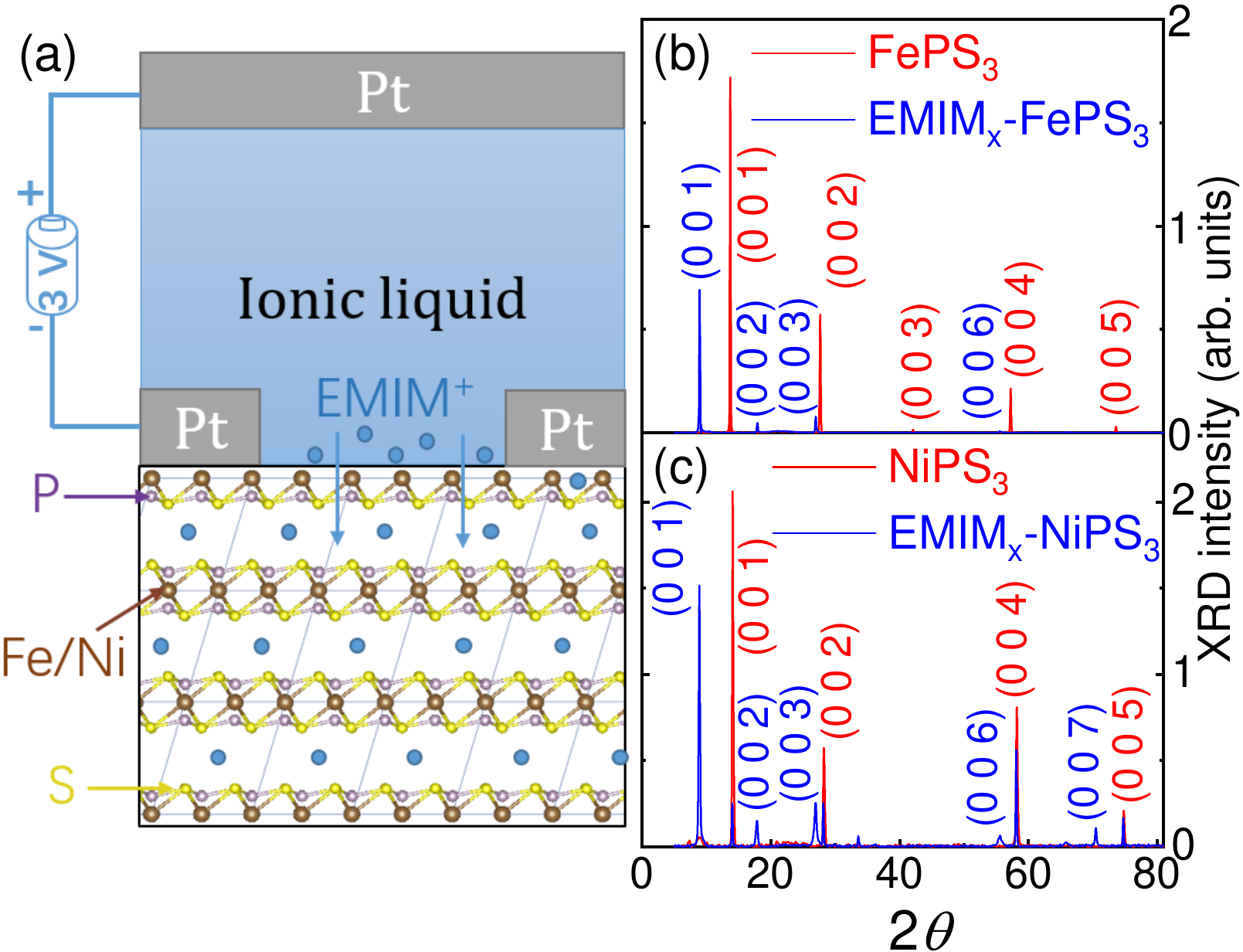}
\caption{\label{xrd}
\textbf{Sample intercalation and structural characterization of FePS$_3$ and NiPS$_3$}. (a) Configuration 
for electrochemical  treatment with EMIM-BF$_4$. Two platinum electrodes are placed in the ionic liquid, 
set to a potential difference of 3~V, and the sample is attached to the negative electrode.
(b)~Single-crystal XRD data for FePS$_3$ and EMIM$_x$-FePS$_3$.
(c)~Single-crystal XRD data for NiPS$_3$ and EMIM$_x$-NiPS$_3$.}
\end{figure}

Figure~\ref{xrd}(b) shows single-crystal XRD data for our FePS$_3$ samples before and after 24-hour 
electrochemical treatment with the ionic liquid EMIM-BF$_4$, and Fig.~\ref{xrd}(c) the analogous data 
for NiPS$_3$. We focus on the (0~0~$L$) Bragg peaks in order to extract the interlayer ($c$-axis) 
dimension of the unit cell. Studying the intralayer structure, meaning the $a$ and $b$ parameters, 
requires powder XRD measurements, but it has been found that the grinding process introduces disorder 
within the layers that precludes a meaningful analysis. The positions of the (0~0~$L$) XRD peaks change
appreciably after the intercalation treatment, and the $c$-axis lattice parameters are shown 
and compared in Table~I. The $c$-axis dimension increases from 6.72~\AA~to 10.40~\AA~for FePS$_3$ and 
from 6.64~\AA~to 10.34~\AA~for NiPS$_3$. Such an extremly large (over 50\%) increase in interlayer 
spacing can be expected to have a very strong effect on reducing the dimensionality to the 2D limit. 

Because the sample is attached to the cathode, and because of the large change in $c$-axis parameter, we 
deduce that EMIM$^+$, rather than H$^+$, is intercalated into the materials. A similar observation of 
intercalation by a large organic molecule has also been reported in NiPS$_3$ using a different type of 
ionic liquid~\cite{Mi_AdvFM_2022}. We note that our intercalation in EMIM$_x$-NiPS$_3$ was not complete 
by volume, in that a small portion of the XRD pattern of EMIM$_x$-NiPS$_3$ coincided with that of pristine 
NiPS$_3$ [visible in Fig.~\ref{xrd}(c)]. However, the volume ratio of residual NiPS$_3$ is rather small, 
as we establish later from the absence of detectable signals in the magnetic susceptibility and NMR spectra.

\begin{table}[b]
 	\centering
 	\label{latticeparameters}
 	\caption{Lattice parameters of FePS$_3$, EMIM$_x$-FePS$_3$, NiPS$_3$, and EMIM$_x$-NiPS$_3$, and the estimated doping, $x$. }
 	\vspace{5pt}
 	\begin{tabular}{c|c|c|c|c}
 		\hline
 		 &{FePS$_3$} & $\,$ {EMIM$_x$-FePS$_3$} $\,$ & {NiPS$_3$} & $\,$ {EMIM$_x$-NiPS$_3$} $\,$ \\
 		\hline
 		{$c$~(\AA)} $\,$ & $\,$ {6.723(1)} $\,$ & {10.401(1)} & $\,$ {6.642(1)} $\,$ & {10.343(1)} \\
  		\hline
   		{$x$~} &{ 0 } &{ $27 \pm 3$\% } &{ 0 } &{ $37 \pm 0.4$\%} \\
  		\hline
 	\end{tabular}
 \end{table}

The mass of the FePS$_3$ crystal increased from $0.18 \pm 0.01$~mg before to $0.21 \pm 0.01$~mg after 
intercalation. From this we estimate the intercalant concentration and the charge doping level to be 
$x = 27 \pm 3\%$, given the molecular mass of 183~g/mol for FePS$_3$ and 111~g/mol for EMIM$^+$.
For the NiPS$_3$ crystal, the mass changed from $3.86 \pm 0.01$~mg to $4.71 \pm 0.01$~mg, and thus we 
extracted the doping level as $x = 37 \pm 0.4\%$, using the molecular mass of 186~g/mol for 
NiPS$_3$. Our NMR studies revealed in addition the presence of F$^{-}$ ions in the intercalated samples, 
but at an unknown small concentration (data not shown), which suggests that the actual doping $x$ may 
vary with position across the samples.

\section{\label{sfeps3}Experimental data on F\lowercase{e}PS$_3$ and EMIM$_{\lowercase{x}}$-F\lowercase{e}PS$_3$  \label{sec:ED}}

\subsection{\label{sfeps3chi}Magnetic susceptibility}

\begin{figure}[t]
\includegraphics[width=8.6cm]{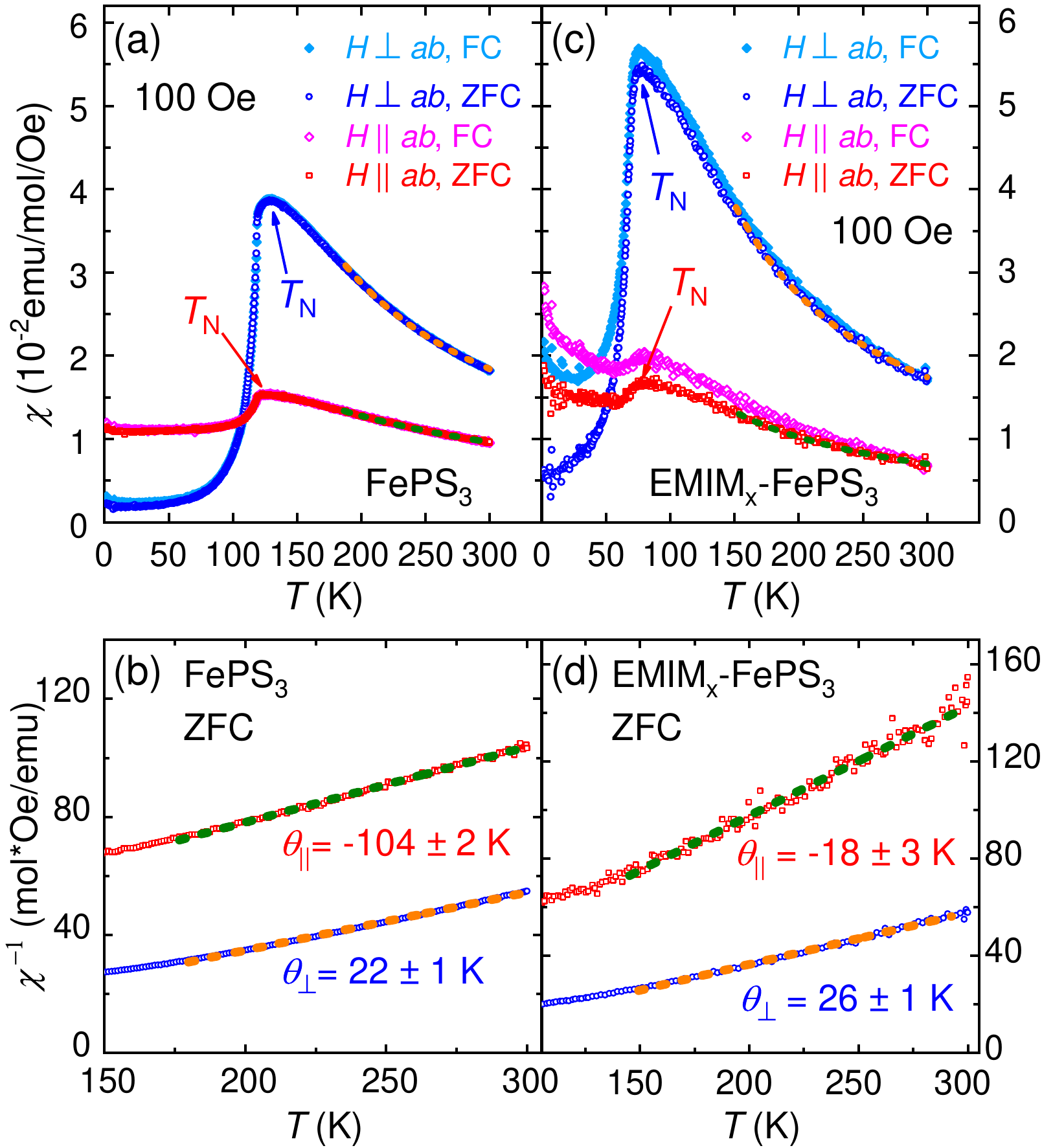}
\caption{\label{feps3sus} \textbf{Magnetic susceptibility of FePS$_3$ and EMIM$_x$-FePS$_3$.} $\chi(T)$ 
measured in a field of 100~Oe applied parallel and perpendicular to the $ab$ plane, under ZFC and FC conditions.
(a) Susceptibility of FePS$_3$ shown as a function of temperature.
(b) $1/\chi (T)$ of FePS$_3$ in the high-temperature PM phase.
(c) Susceptibility of EMIM$_x$-FePS$_3$ shown as a function of temperature.
(d) $1/\chi (T)$ of EMIM$_x$-FePS$_3$ in the high-temperature PM phase.
$T_N$ in panels (a) and (c) labels the AFM transition temperature, which is located at the peak of $\chi(T)$.
Solid lines in panels (b) and (d) are linear fits to the data made over the temperature range shown, with 
$\theta$ obtained from the intercept on the $x$ axis (i.e.~at $y = 0$).}
\end{figure}

Our susceptibility measurements were conducted on single crystals of FePS$_3$ and EMIM$_x$-FePS$_3$ 
with a small field applied parallel and perpendicular to the crystalline $ab$ plane, and data under 
FC and ZFC conditions are shown in Fig.~\ref{feps3sus}. In both pure and intercalated FePS$_3$, 
$\chi(T)$ is much bigger at temperatures above $T_N$ when the field is perpendicular to the $ab$ 
plane than when it lies in the $ab$ plane, but much smaller below $T_N$. Such an anisotropy should 
indicate the Ising nature of the system, which is attributed to combination of trigonal 
distortion and spin-orbit coupling~\cite{Chatterjee_PRB_1995,Pramana_1994,Joy_PRB_1992,Peter_PRB_2023}. 
With $H \perp ab$, $\chi$ does not reach zero in the zero-temperature limit, which may be the consequence 
of a small field misalignment.

On cooling in FePS$_3$ [Fig.~\ref{feps3sus}(a)], $\chi(T)$ exhibits a peak followed by a sharp drop at 120~K 
in both field orientations, indicating the N\'{e}el transition at $T_N \simeq 120$~K as found in previous
reports~\cite{Joy_PRB_1992,Flem_JPCS_1982,Okuda_1983}. There is no appreciable hysteresis with FC and ZFC
conditions, which supports a second-order AFM transition, in contrast to earlier reports that this transition 
may be of first order~\cite{Jernberg_JM_1984,Ferloni_ThermochimicaActa_1989}. Turning to EMIM$_x$-FePS$_3$, 
$T_N$ taken from the peak position in $\chi(T)$ is reduced to 80~K by intercalation [Fig.~\ref{feps3sus}(c)], 
which is a value still 2/3 of that in the pristine sample. These values of $T_N$ will be verified by the NMR 
measurements to follow, and already contain one of our primary results, that there is little to no connection 
between $T_N$ and a hypothetical interlayer coupling. The FC and ZFC data for EMIM$_x$-FePS$_3$ already differ 
at temperatures above $T_N$, but this difference is affected only minimally by the onset of magnetic order at 
$T_N$; thus we believe that the difference between the FC and ZFC datasets is not an intrinsic effect, but an 
extrinsic one caused by impurities that enter the sample during chemical intercalation.

In all cases shown in Fig.~\ref{feps3sus}, $\chi(T)$ above $T_N$ can be well fitted by the Curie-Weiss (CW) 
form, $\chi = \dfrac{N \mu_0 \mu_{\rm eff}^2}{3 k_B (T - \theta)}$, where $\mu_{\rm eff}$ is the effective 
paramagnetic (PM) moment and $\theta$ is the CW temperature. In FePS$_3$ [Fig.~\ref{feps3sus}(b)], the 
CW temperatures are found to be $\theta_\perp = 22 \pm 1$~K and $\theta_\parallel = - 104 \pm 2$~K for 
the two field orientations. In our convention, the positive $\theta_\perp$ indicates the presence of FM 
interactions, as one may expect from the zig-zag ground-state order; nevertheless, this ordered state is 
dominated by AFM interactions and by the Ising anisotropy, so our result reflects the magnetic frustration 
in the system. The CW fit to the perpendicular-field data returns a PM moment $\mu_{\rm eff} = 6.4 \pm 
0.2 \mu_B$/Fe, which is close to the literature value~\cite{Joy_PRB_1992}; as noted in early studies, such 
a large $\mu_{\rm eff}$ for divalent Fe$^{2+}$ ions can be explained only by taking the spin-orbit coupling 
into account~\cite{Joy_PRB_1992}.

For EMIM$_x$-FePS$_3$, the susceptibility shows the same type of anisotropy as the pristine sample 
[Fig.~\ref{feps3sus}(c)]. The PM moments we extract, $6.0 \pm 0.2 \mu_B$/Fe, are largely unaltered, 
indicating that intercalation causes no significant changes to the ionic properties. The CW temperatures 
we obtain, $\theta_\perp = 26 \pm 1$~K and $\theta_\parallel = - 18 \pm 3$~K [Fig.~\ref{feps3sus}(d)], 
suggest that no dramatic changes take place in the signs or energy scales of the couplings, but offer no 
straightforward interpretation as alterations to the dominant AFM or frustrating FM interactions. 
Because the susceptibility measurements contain significant impurity contributions whose subtraction is 
difficult to benchmark, we turn next from a bulk to a microscopic probe in order to gain deeper insight 
into the magnetic properties after intercalation.

\subsection{\label{sfeps3spec}NMR spectra}

\begin{figure}[t]
\includegraphics[width=8.6cm]{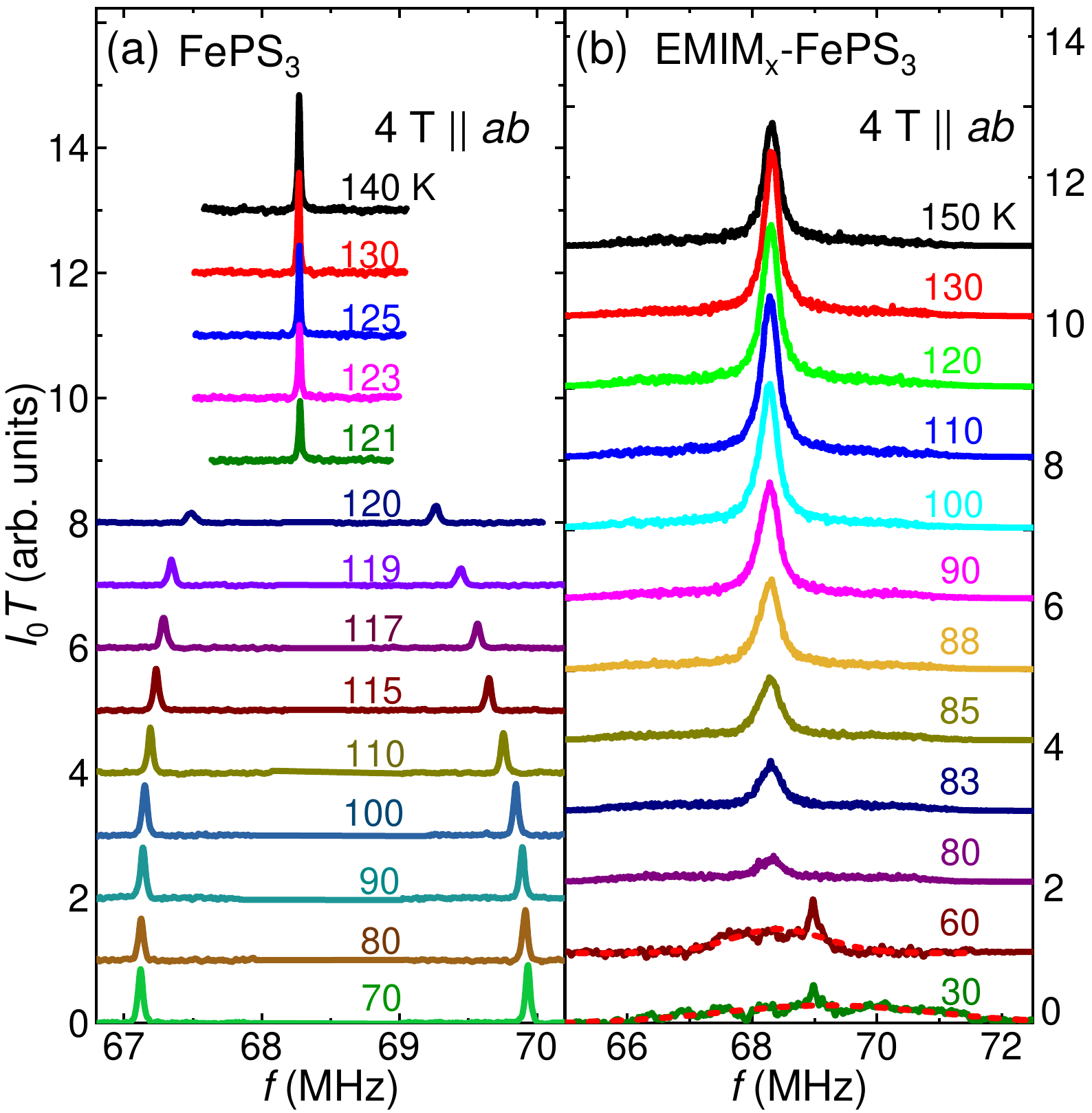}
\caption{\label{feps3spec}
\textbf{$^{31}$P NMR spectra of FePS$_3$ and EMIM$_x$-FePS$_3$.} (a) Spectra of FePS$_3$ over a range 
of temperatures above and below $T_N$, taken at a fixed in-plane field of 4~T. Successive datasets are 
offset vertically for clarity. (b) Spectra of EMIM$_x$-FePS$_3$ taken under the same conditions. The 
spectral intensities at 30 and 60~K are multiplied by a factor of two; dashed lines represent Gaussian 
fits to the data.}
\end{figure}

\begin{figure}[t]
\includegraphics[width=8.6cm]{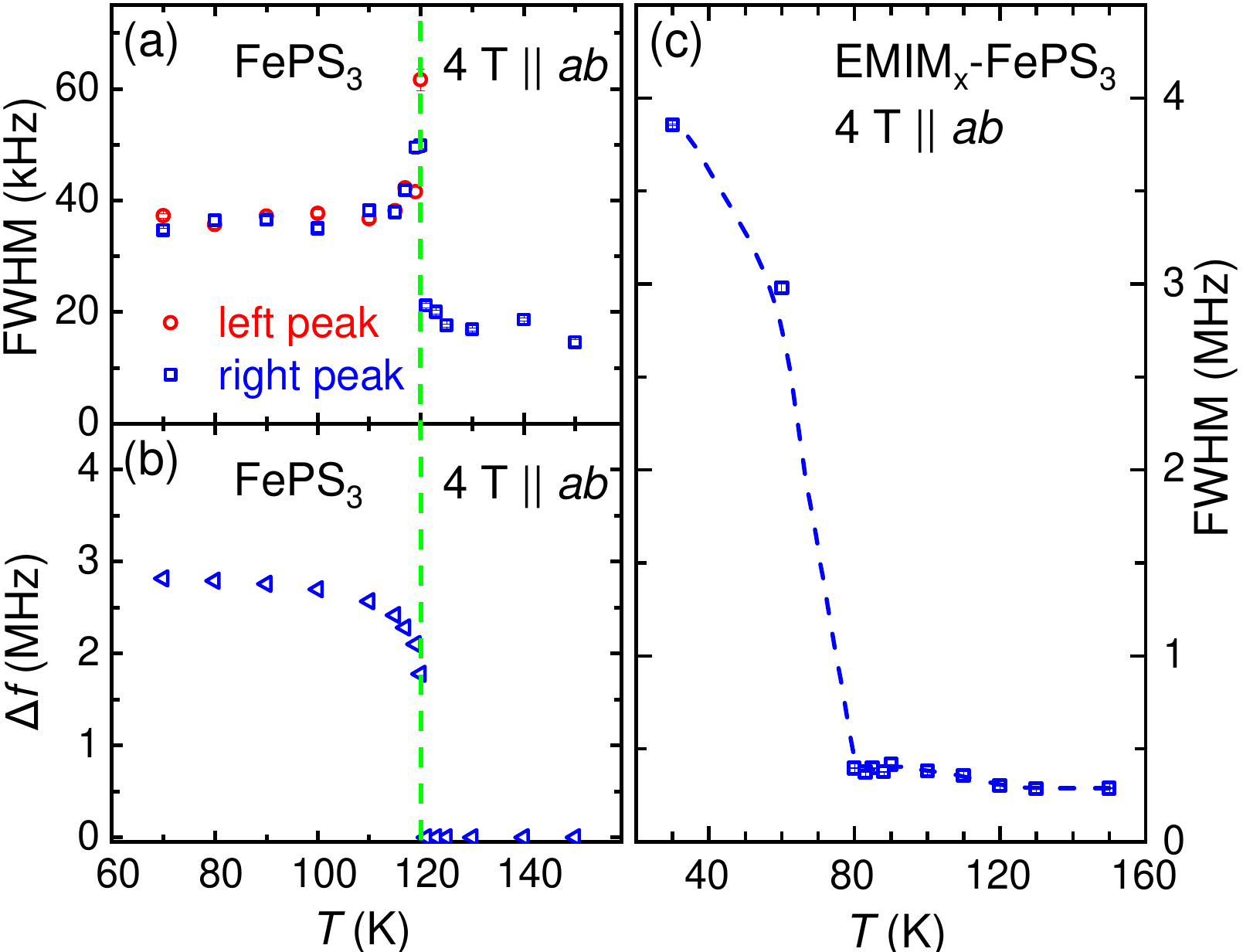}
\caption{\label{feps3anysis} \textbf{NMR line widths and splittings in FePS$_3$ and EMIM$_x$-FePS$_3$.}
(a) FWHM of the primary spectral peak, or peaks, in FePS$_3$ [Fig.~\ref{feps3spec}(a)] shown as a 
function of temperature. (b) Splitting, $\Delta f$, of NMR peaks in FePS$_3$ as a function of temperature.
(c) FWHM of the broad NMR spectrum of EMIM$_x$-FePS$_3$ [Fig.~\ref{feps3spec}(b)] shown as a function 
of temperature.}
\end{figure}

We measured $^{31}$P NMR spectra for FePS$_3$ and EMIM$_x$-FePS$_3$ over a wide range of temperatures on 
both sides of $T_N$, and show our results in Fig.~\ref{feps3spec}. In FePS$_3$, the single NMR peak observed
at temperatures above 120~K splits into two nearly symmetrical peaks, presenting a clear signature of AFM 
ordering at $T_N$ [Fig.~\ref{feps3spec}(a)]. We note here that the field was applied in the $ab$ plane, 
where from Fig.~\ref{structure}(b) one expects the $^{31}$P nucleus to experience hyperfine fields from the 
two different sites that are in opposite directions relative to the external field. We then fitted the peak 
or peaks in each NMR spectrum with Gaussian profiles to extract the full width at half-maximum height 
(FWHM) and the peak splitting below $T_N$. As Fig.~\ref{feps3anysis}(a) shows, the single NMR peak above 
120~K broadens rapidly at $T_N$, from approximately 18 to over 50~kHz where it splits. The line splitting, 
${\Delta}f$, approaches 2.8~MHz at low temperatures and, as shown in Fig.~\ref{feps3anysis}(b), from its 
shape can serve as an order parameter for the AFM phase.

For EMIM$_x$-FePS$_3$, the single NMR peak remains rather narrow when cooled down to 80~K 
[Fig.~\ref{feps3spec}(b)]. Here it shows no evidence of a line splitting, but instead undergoes 
a large increase in line width [Fig.~\ref{feps3anysis}(c)]. This broadening can be taken to 
indicate the onset of magnetic order at $T_N \simeq 80$~K, consistent with the susceptibility data 
of Fig.~\ref{feps3sus}, and Gaussian peak fits show the FWHM broadening from 300~kHz above $T_N$ 
to 3~MHz below $T_N$. We draw attention to the fact that the low-temperature FWHM in EMIM$_x$-FePS$_3$ 
has the same order as ${\Delta}f$ in FePS$_3$, which suggests a distribution of ordered moments in the 
system. Because we observe only one peak, we propose that these results reflect an incommensurate 
magnetic order in the 2D layers, with a systematic distribution of local-moment magnitudes and/or 
directions, as opposed to a multi-domain structure. Such an incommensurate order could arise from 
rather small alterations of the intralayer interactions in FePS$_3$, whose pre-intercalation 
compromise configuration is the Ising zig-zag order shown in Fig.~\ref{structure}(b).

\subsection{\label{sfeps3slrr}Spin-lattice relaxation rates and spin gaps}

\begin{figure}[t]
\includegraphics[width=8.6cm]{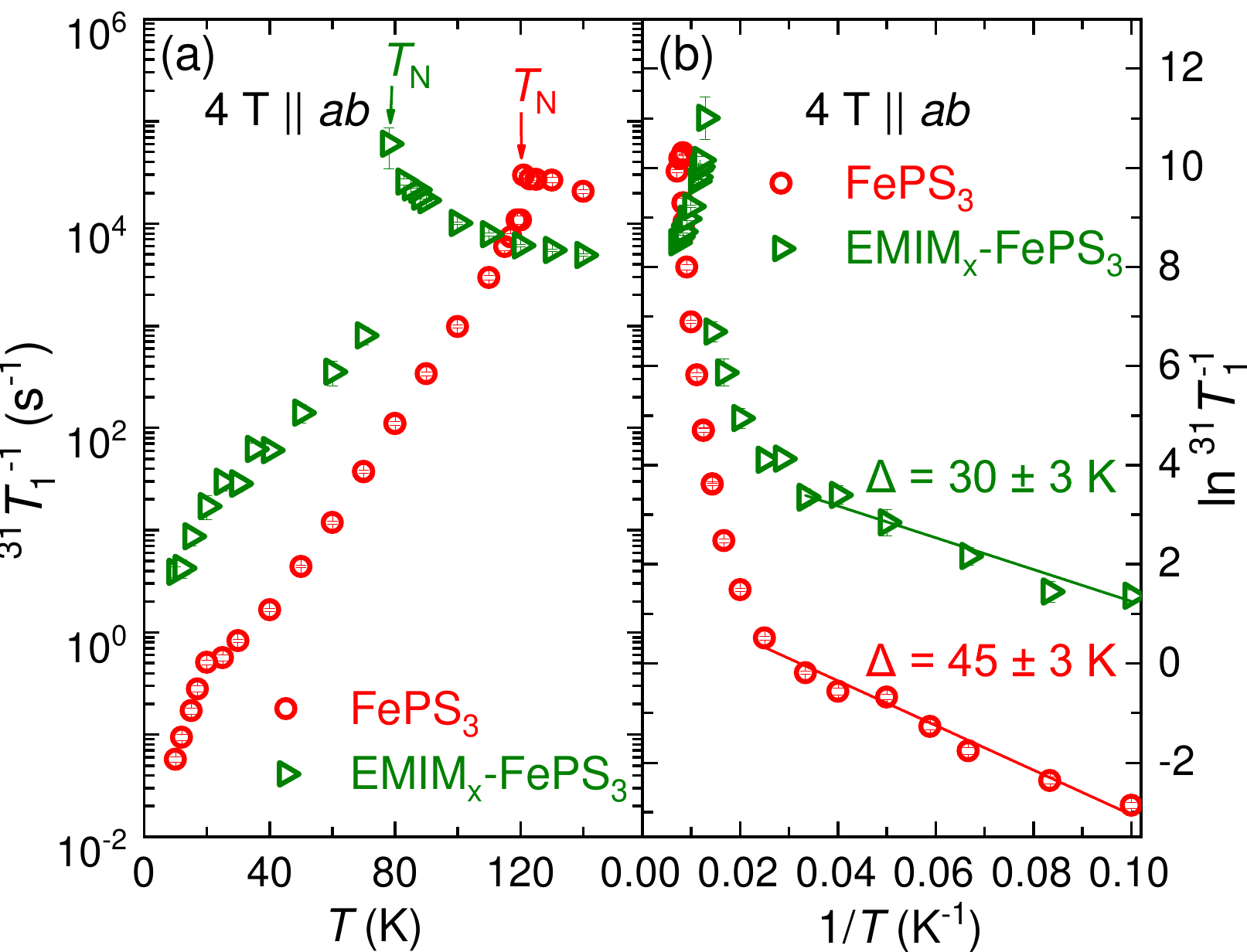}
\caption{\label{feps3T1} \textbf{Spin-lattice relaxation rates of FePS$_3$ and EMIM$_x$-FePS$_3$.}
(a) $1/^{31}T_1$ shown as a function of temperature, measured with an in-plane field of 4~T. 
(b) $1/^{31}T_1$ shown in semilog form as a function of $1/T$. The straight solid lines represent 
linear fits to the data in the low-temperature regime. $\Delta$ is a gap value extracted from the 
gradient of each line.}
\end{figure}

The spin-lattice relaxation rates, $1/^{31}T_1$, of pure and intercalated FePS$_3$ are shown in 
Fig.~\ref{feps3T1}. FePS$_3$ exhibits the characteristic behavior on decreasing temperature that 
$1/^{31}T_1$ first increases to a maximum at $T = 120$~K, which reflects the AFM transition at $T_N$, 
followed by a rapid drop [Fig.~\ref{feps3T1}(a)]. Far below $T_N$, the exponential fall suggests the 
form $1/^{31}T_1 \propto e^{-{\Delta}/k_B T}$, where $\Delta$ is the spin gap expected in an ordered 
Ising magnet. Plotting $\ln (1/^{31}T_1)$ as a function of $1/T$ reveals that the data below 50~K 
follow a straight line [Fig.~\ref{feps3T1}(b)], whose gradient gives a gap $\Delta = 45 \pm 3$~K.
Here we comment that inelastic neutron scattering (INS) measurements of the anisotropic spin-wave 
spectra have reported a gap of 15~meV~\cite{Wildes_PRB_2016}, which is significantly larger than our 
NMR value. Because the NMR $1/T_1$ is a very sensitive, low-energy local probe that sums contributions 
from all of momentum space, it is possible that our results reveal a process missed by both triple-axis 
and time-of-flight INS for reasons of energy resolution or reciprocal-space coverage; however, the 
higher temperatures at which we obtained an NMR signal also leave open the possibility of probing a 
finite-energy transition, and we hope that future studies may resolve this apparent contradiction.

Similar magnetic ordering and spin-gap behavior are observed in EMIM$_x$-FePS$_3$, as 
Fig.~\ref{feps3T1}(a) shows. Here $T_N$ is determined quite precisely at 78~K from the sharp peak in 
$1/^{31}T_1$. Analysis of the low-temperature response, shown in Fig.~\ref{feps3T1}(b), gives the 
equivalent spin gap as $\Delta = 30 \pm 3$~K. This suppression of the gap upon intercalation scales 
linearly with the suppression of $T_N$, which would be consistent with a straightforward 
reduction of the relevant intralayer energies.

\section{\label{snips3}Experimental data on N\lowercase{i}PS$_3$ and EMIM$_{\lowercase{x}}$-N\lowercase{i}PS$_3$  \label{sec:don}}

\subsection{\label{snips3chi}Magnetic susceptibility}

\begin{figure}[t]
\includegraphics[width=8.6cm]{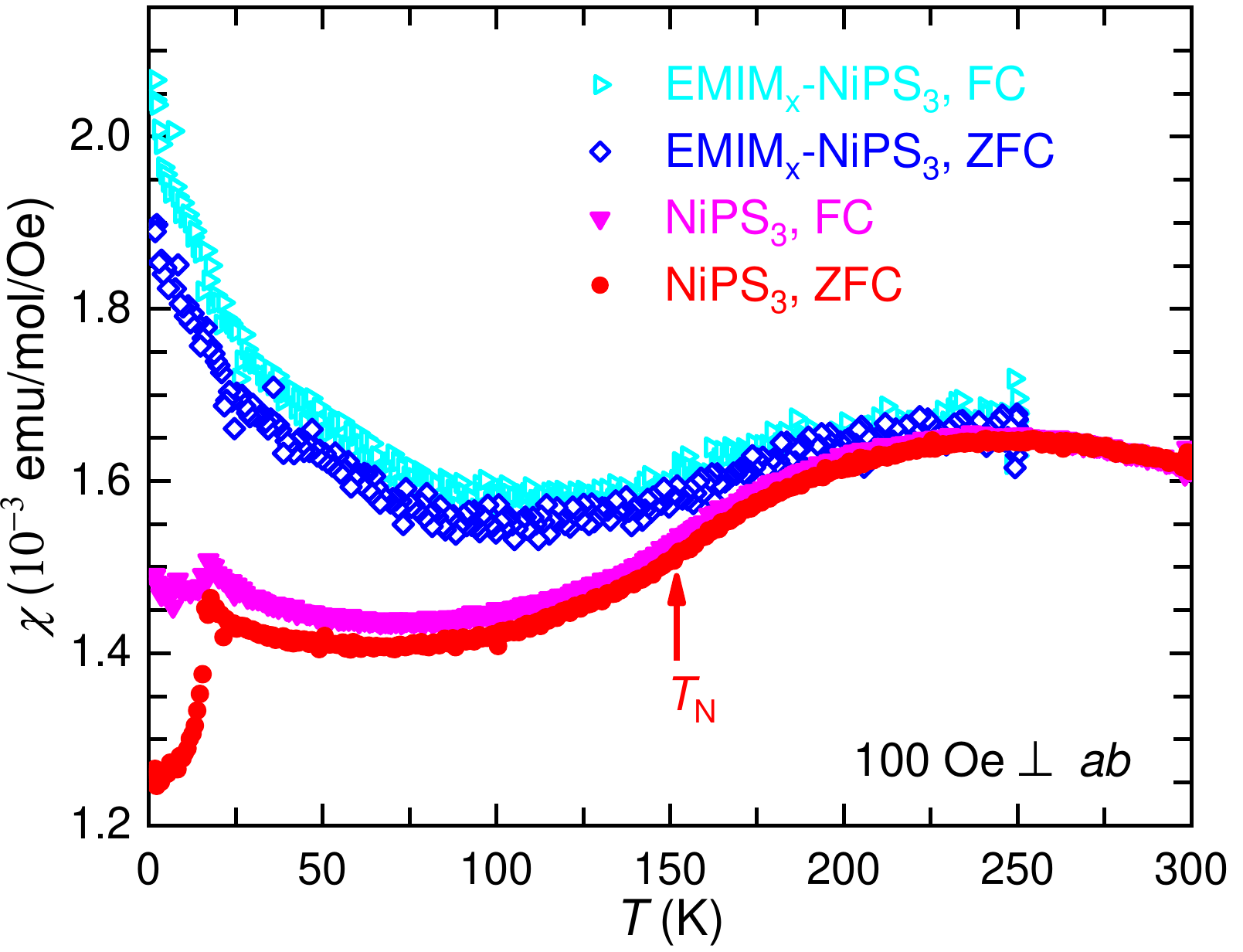}
\caption{\label{nips3sus} \textbf{Magnetic susceptibility of NiPS$_3$ and EMIM$_x$-NiPS$_3$.} 
$\chi (T)$ measured in an out-of-plane field of 100~Oe under FC and ZFC conditions. }
\end{figure}

\begin{figure}[t]
\includegraphics[width=8.6cm]{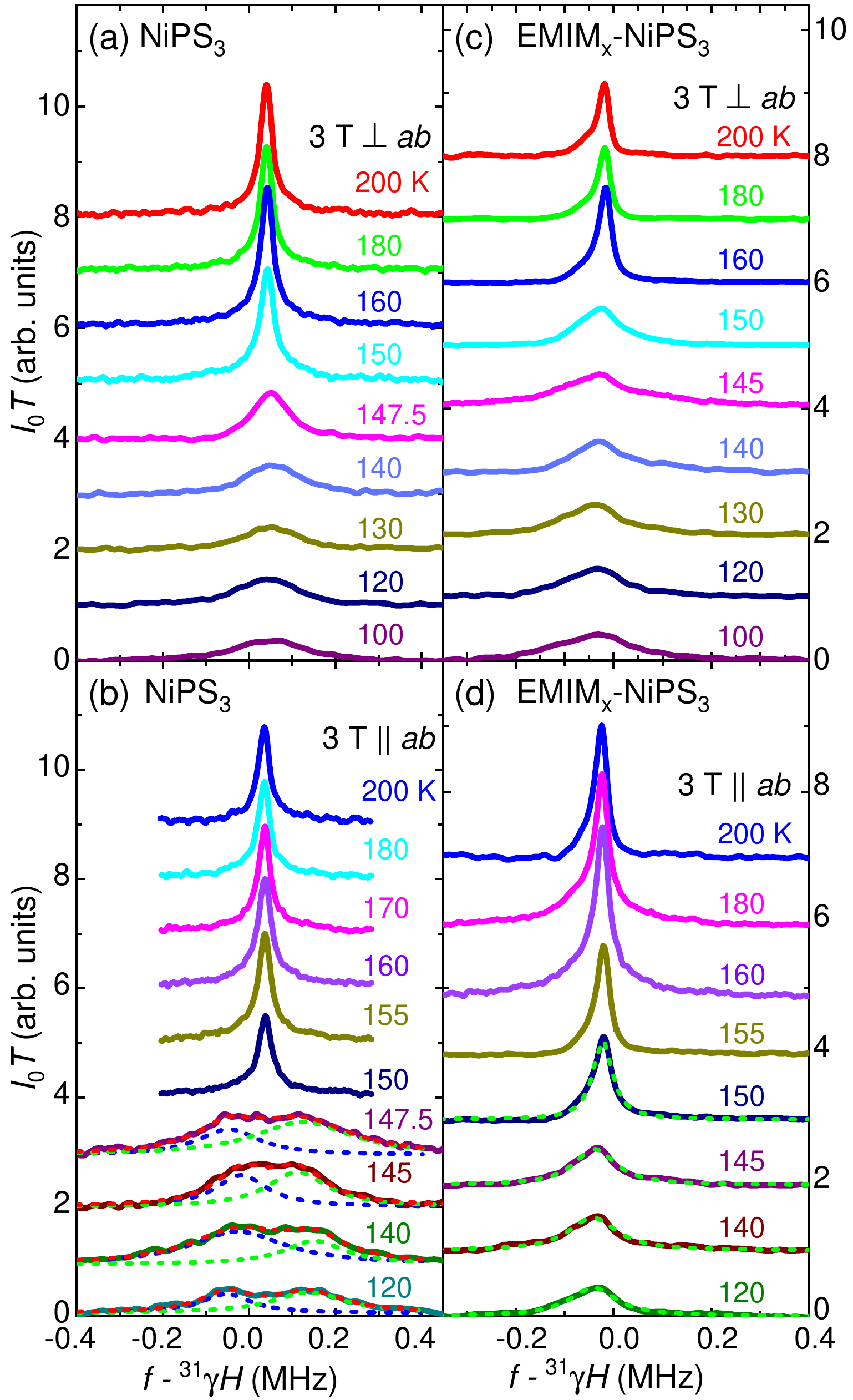}
\caption{\label{nips3spec} \textbf{$^{31}$P NMR spectra of NiPS$_3$ and EMIM$_x$-NiPS$_3$.}
(a) Spectra of NiPS$_3$ measured in an out-of-plane field of 3~T over a wide range of temperatures 
above and below $T_N$. Successive datasets are offset vertically for clarity. (b) As in panel (a) 
with an in-plane field. The low-temperature spectra are fitted to double-Lorentzian functions.
(c) and (d) Spectra of EMIM$_x$-NiPS$_3$ measured under the same conditions of field and temperature.}
\end{figure}

The magnetic susceptibilities of NiPS$_3$ and EMIM$_x$-NiPS$_3$ are shown in Fig.~\ref{nips3sus}. For both 
compounds, the peak in $\chi (T)$ occurs around 250~K, which makes it impossible to reach a high-temperature 
regime in which to extract the CW behavior. Such high intralayer energy scales are found by high-precision 
INS measurements of the spin dynamics of NiPS$_3$~\cite{Wildes_PRB_2022}, and the broad peak in $\chi$ 
indicates the development of 2D AFM correlations. We note that the AFM ordering transition is essential invisible 
in $\chi(T)$, and we will show by NMR that $T_N = 148$~K (below). Detailed measurements on NiPS$_3$ showed 
that the susceptibility is isotropic above $T_N$ and anisotropic below it~\cite{Wildes_PRB_2015}. The most 
dramatic feature in our measurements (Fig.~\ref{nips3sus}) is the onset of a large difference between the 
FC and ZFC data below 25~K, which may be caused by quenched disorder or a change of magnetic structure.

For EMIM$_x$-NiPS$_3$, $\chi (T)$ is remarkably similar to the pristine compound, with only a small deviation 
setting in at temperatures below 200~K. This minimal alteration despite the massive increase in interlayer 
spacing is a strong indication that all the magnetic properties of NiPS$_3$ are intrinsically 2D. For the 
intercalated system, a difference between the ZFC and FC data develops gradually with decreasing temperature, 
unaffected by the onset of magnetic order (shown below by NMR), and we believe this to be a consequence of 
impurities introduced by the intercalation, similar to the situation in EMIM$_x$-FePS$_3$ (Fig.~\ref{feps3sus}).

\subsection{\label{snips3spec}NMR spectra}

\begin{figure}[t]
\includegraphics[width=8.6cm]{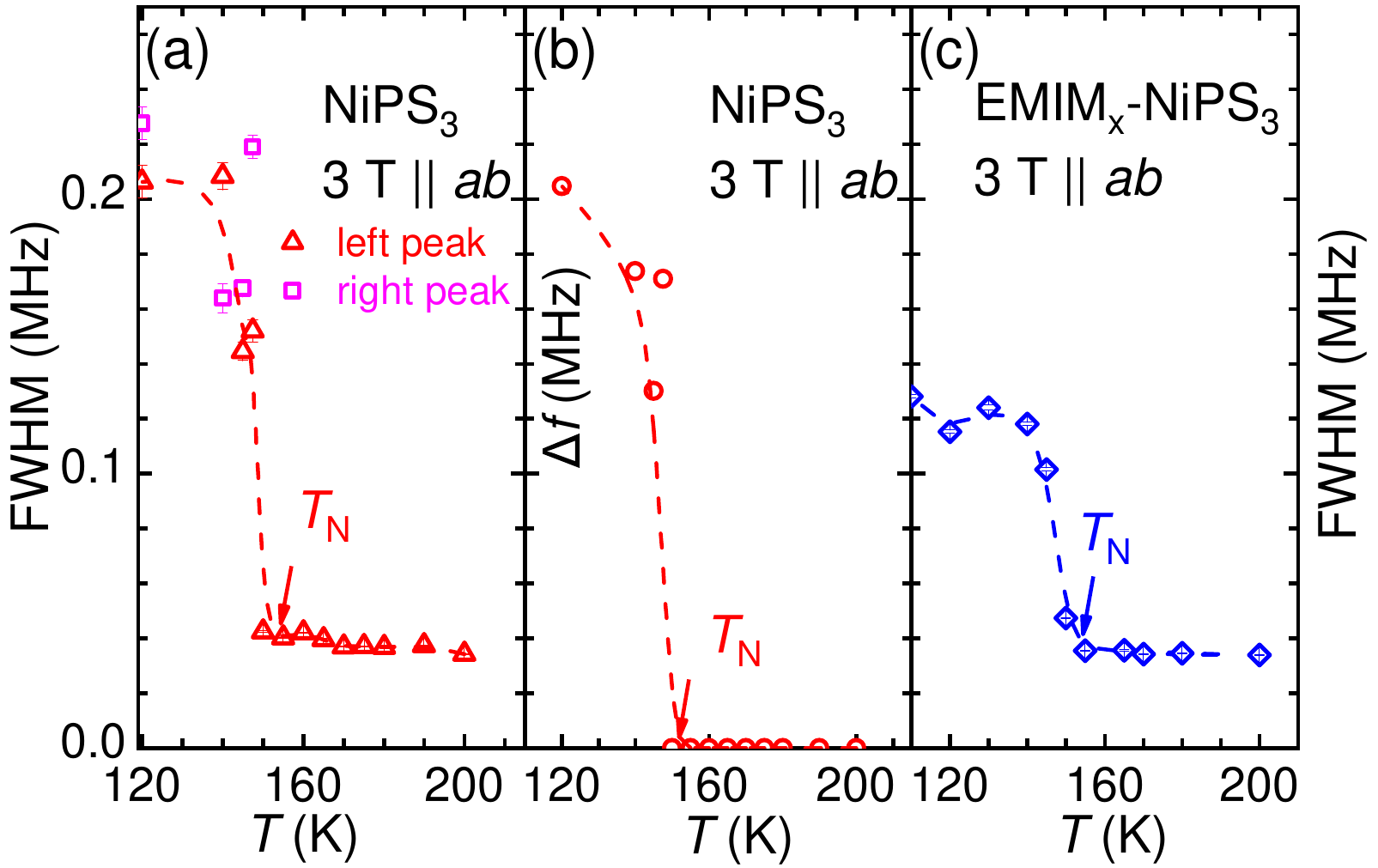}
\caption{\label{nips3specanalysis} \textbf{NMR line widths and splittings in NiPS$_3$ and EMIM$_x$-NiPS$_3$.}
(a) FWHM of the primary spectral peak, or peaks, in NiPS$_3$ [Figs.~\ref{nips3spec}(b)] shown as a 
function of temperature. (b) Splitting, $\Delta f$, of NMR peaks in NiPS$_3$ as a function of temperature.
(c) FWHM of the broad NMR spectrum of EMIM$_x$-NiPS$_3$ [Figs.~\ref{nips3spec}(d)] shown as a function 
of temperature. The arrows mark the transition temperatures, $T_N$ and $T^*$, extracted for the two compounds.}
\end{figure}

Figure~\ref{nips3spec} shows the $^{31}$P NMR spectra of NiPS$_3$ and EMIM$_x$-NiPS$_3$ over a wide range of 
temperature for both field orientations. For NiPS$_3$ in an out-of-plane field, the spectra above 148~K have 
a single peak characteristic of the PM phase, which broadens significantly on further cooling from 148~K to 
140~K [Fig.~\ref{nips3spec}(a)]. By contrast, the double-peak feature developing at all temperatures below 
148~K in an in-plane field demonstrates clearly the onset of AFM order and fixes $T_N = 148 \pm 2$~K. In this 
case, Lorentzian functions gave a slightly better fit to the spectra, although this choice made little difference for 
the purpose of extracting the FWHM of each peak and the line splitting, $\Delta f$, below $T_N$ [shown in 
Figs.~\ref{nips3specanalysis}(a) and \ref{nips3specanalysis}(b)]. The FWHM is about 40~kHz at temperatures 
above $T_N$ and saturates around 0.2~MHz far below $T_N$, while ${\Delta}f$ also tends to level off around 
0.2~MHz at low temperatures in the ordered state.

For EMIM$_x$-NiPS$_3$ in both field orientations, the NMR peaks also broaden significantly towards low 
temperatures [Fig.~\ref{nips3spec}(c,d)]. However, we were not able to resolve any line splitting, even in 
an in-plane field. Again we used Lorentzian fits, to the peaks obtained in in-plane fields, in order to 
extract the FWHM shown in Fig.~\ref{nips3specanalysis}(c). The sudden increase in FWHM occurring 
at 145~K, which we label $T^*$, is characteristic of a static magnetic order. However, the absence of any
line splitting at any lower temperatures suggests that this order could be of a type different from that 
in pristine NiPS$_3$.

\subsection{\label{snips3slrr}Spin-lattice relaxation rates}

\begin{figure}[t]
\includegraphics[width=8.6cm]{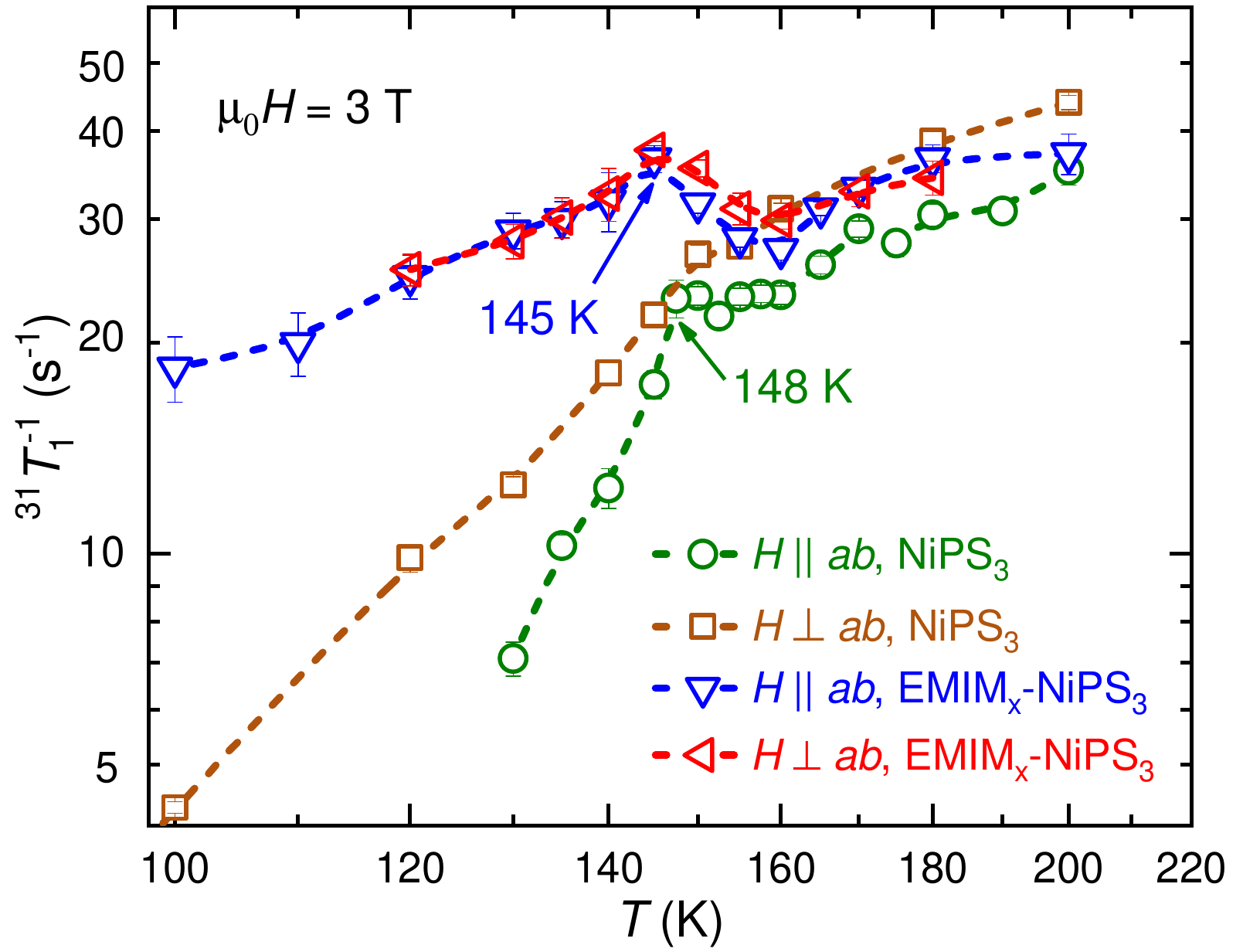}
\caption{\label{nips3T1} \textbf{Spin-lattice relaxation rates of NiPS$_3$ and EMIM$_x$-NiPS$_3$.}
$1/^{31}T_1$ shown as a function of temperature, measured with in- and out-of-plane fields of 3~T.} 
\end{figure}

The spin-lattice relaxation rates of NiPS$_3$ and EMIM$_x$-NiPS$_3$, measured in both field orientations, 
are presented in Fig.~\ref{nips3T1}. We comment that it was not possible to obtain a reliable 
$1/^{31}T_1$ signal at temperatures significantly below 100~K, in contrast to the situation in FePS$_3$ 
(Fig.~\ref{feps3T1}), but our data characterize well the magnetic transitions in both materials, which 
are clearly evident from the kinks in $1/^{31}T_1$. For NiPS$_3$ in both orientations, the rapid drop 
in $1/^{31}T_1$ determines precisely the AFM ordering temperature $T_N = 148$~K, a value slightly lower 
than that obtained from early susceptibility data~\cite{Flem_JPCS_1982,Joy_PRB_1992,Ziolo_JAP_1988}. 
Below $T_N$, $1/^{31}T_1$ for an in-plane field drops significantly faster than with an out-of-plane 
field, which is consistent with XY magnetic order~\cite{Rosen_PRB_1994,Kim_NC_2019,Jana_PRB_2023} in 
that the gapless (Goldstone) modes should be preserved only when the field is applied out of the plane.

For EMIM$_x$-NiPS$_3$, our $1/^{31}T_1$ data for the two field orientations are barely distinguishable 
above $T^*$, where they decrease with decreasing temperature from 200~K to 160~K and then increase over 
the range from 160~K to 145~K. Below this they drop more rapidly, allowing us to fix the putative magnetic 
ordering temperature as $T^* = 145$~K. Below $T^*$, the relaxation rates under in- and out-of-plane fields 
are truly indistinguishable, presenting the first of two distinctive contrasts between NiPS$_3$ and 
EMIM$_x$-NiPS$_3$. Such isotropic behavior suggests that intercalation in this system genuinely alters 
the intralayer magnetic interactions to a Heisenberg form. Second, although $T^*$ is reduced by only 3~K 
on intercalation, $1/^{31}T_1$ at $T = 100$~K (far below $T^*$) is enhanced by a factor of five. This 
apparent strong enhancement of low-energy spin fluctuations points once again to the possibility of a 
magnetically ordered state different from that of pure NiPS$_3$ [Fig.~\ref{structure}(b)], candidates 
for which would include an incommensurate order. Further measurements are therefore required to 
investigate the magnetic structure of the intercalated system.

\begin{table}[t]
	\centering
	\label{exchanges}
	\caption{Magnetic interactions of FePS$_3$ and NiPS$_3$~\cite{Wildes_PRB_2016,Wildes_PRB_2022}; 
                 $J_i$ indicates a Heisenberg interaction between neighboring ions of differing separations 
                 [Fig.~\ref{structure}(b)] and $D_z$ the leading magnetic anisotropy fitted by INS.}
	\vspace{5pt}
	\begin{tabular}{c|c|c|c|c|c}
		\hline
		{} & {$J_1$ (meV)} & {$J_2$ (meV)} & {$J_3$ (meV)} & {$J^\prime$ (meV)} & {$D_z$ (meV)}\\
		\hline
		{FePS$_3$} &{$-1.46$} &{0.04} &{0.96} & {0.0073}& {$-2.66$}\\
		\hline
		{NiPS$_3$} & {$-2.6$} &{0.2} & {13.5} & {$-0.3$} & {0.21}\\
		\hline
	
	\end{tabular}
\end{table}

\section{\label{sdiscu}Discussion}

Our susceptibility data show that the intercalation of FePS$_3$ and NiPS$_3$ does not cause any depletion 
of the available magnetic moments. Thus the changes we observe, specifically the reduction of $T_N$ in 
EMIM$_x$-FePS$_3$ and the isotropic spin response of EMIM$_x$-NiPS$_3$, should be attributed to changes of 
the magnetic interactions. To discuss these, in Table~II we list the values of the magnetic interactions 
deduced by INS experiments for FePS$_3$ \cite{Wildes_PRB_2016} and NiPS$_3$ \cite{Wildes_PRB_2022}. The 
authors of both studies adopted a minimal model consisting of the three intralayer Heisenberg interactions 
illustrated in Fig.~\ref{structure}(b), an interlayer Heisenberg interaction ($J'$), and single-ion 
anisotropy terms, of which only one was found to be significant. The $J'$ term was found to be very small 
in FePS$_3$ but potentially significant in NiPS$_3$. The $J_3$ term, for third-neighbor $M^{2+}$ ions, was 
found to be completely dominant in NiPS$_3$ and to play an essential role in establishing the zig-zag order 
of FePS$_3$.

Considering the increase of the $c$-axis lattice parameter that we achieve, in excess of 50\%, it is natural 
to expect that the interlayer couplings should be reduced dramatically. Thus our primary conclusion is that 
the $J'$ terms have no role in the magnetism of either material, where 2D ordering remains well established 
in both in EMIM$_x$-FePS$_3$ and in EMIM$_x$-NiPS$_3$. In the light of theoretical studies demonstrating that 
quasi-2D magnets with very small $J'$ may nevertheless have significant $T_N$ values \cite{Yasuda_PRL_2018}, 
this result was not fully clear for FePS$_3$ (Table II). Our study demonstrates that it is also true for 
NiPS$_3$, despite the 3~K interlayer coupling scale extracted in Ref.~\cite{Wildes_PRB_2022}.

It is then important to understand how the intralayer magnetic interactions are modified by intercalation. 
In principle, the addition of EMIM$^{+}$ should lead to electron doping in both materials, with the 
electrons presumably located on the Fe or P ions, and reaching a doping around 1/3~e$^-$/f.u. (0.27 in 
EMIM$_x$-FePS$_3$, 0.37 in EMIM$_x$-NiPS$_3$). Considering the unchanged paramagnetic moments we extracted 
for EMIM$^{+}$-FePS$_3$ and the unchanged magnetic properties of EMIM$^{+}$-NiPS$_3$, a change of the $M^{2+}$ 
valence seems to be excluded. Thus we propose that the electronic doping affects the P sites, by transforming 
the less favorable P$^{4+}$ ions~\cite{Guo_JMCA_2019} to P$^{3+}$. With the P ions located at the centers 
of the $M$ hexagons, we expect that this doping should be very effective in modifying the intralayer 
superexchange interactions, particularly the strong $J_3$ terms. We remark in this context that we found 
the intercalated samples to remain highly insulating, which is a prerequisite to regard the doped charges 
as localized, such that they can affect the magnetic interactions rather than drive a metallic transition.

Considering FePS$_3$ in more detail, the $T_N$ we measure for bulk EMIM$_x$-FePS$_3$ is lower 
than the value $T_N = 104$~K obtained for monolayer FePS$_3$~\cite{Wang_2DMaterial_2016}. Thus our 
intercalation is more than a simple isolation of the 2D units, and its effects on the competing intralayer 
interactions require more detailed microscopic modelling. In fitting their INS results, the authors of 
Ref.~\cite{Wildes_PRB_2016} assumed an extremely minimal model where all of the anisotropy was ascribed 
to one single-ion term, and further anisotropic interactions such as Dzyaloshinskii-Moriya or Ising-type 
spin-spin interactions were neglected. The modelling of these terms is a complex task that was approached 
only recently by correlated density functional theory (DFT) methods \cite{Peter_PRB_2023}, while further 
experimental information on the magnetic anisotropy has been obtained by X-ray photoelectron microscopy 
\cite{Lee_AEM_2023}. With the ongoing interest in FePS$_3$ as a candidate spintronic material 
\cite{srfp_2016}, and for band engineering by ultrafast coherent light \cite{Mirko_AM_2023}, a deeper 
understanding of the key interaction terms remains a priority. 

In EMIM$_x$-NiPS$_3$, the fact that $T^*$ is only 3~K lower than $T_N$ in NiPS$_3$ demonstrates 
immediately that $J'$ terms are irrelevant. Nevertheless, the absence of line splitting at low temperatures 
in the $^{31}$P spectra of EMIM$_x$-NiPS$_3$ [Fig.~\ref{nips3spec}(d)], in contrast to the behavior in 
NiPS$_3$ [Fig.~\ref{nips3spec}(b)], and the enhancement of $1/^{31}T_1$ (Fig.~\ref{nips3T1}), lead us to 
suggest that the ordering at $T^*$ may be of a different type, such as incomplete (short-ranged) or 
incommensurate. Recent theoretical work has indeed proposed that NiPS$_3$ is close to incommensurate 
magnetic order~\cite{Paula_APL_2023}, and the large overlap of the two NMR peaks we observe in 
Fig.~\ref{nips3spec}(b) may support this scenario. Another possibility could be weak anisotropy caused 
by intralayer strain arising from a lattice mismatch with the size of the intercalated molecules. Finally, 
a BKT transition is also possible if the system retains an easy-plane anisotropy, however weak, and in 
the intercalated system it is likely that quenched disorder would help to pin true long-range order from 
the quasi-long-ranged BKT order. We note here that a BKT phase has already been reported in monolayer 
NiPS$_3$~\cite{Hu_PRB_2023}.

In summary, we performed electrochemical treatment of FePS$_3$ and NiPS$_3$ with the ionic liquid 
EMIM-BF$_4$, obtaining a significant intercalation of the large EMIM$^+$ ions into both compounds. 
The most important consequence of intercalation is a large and uniform increase of the $c$-axis 
lattice parameters, which unavoidably causes a strong reduction of interlayer couplings. The fact 
that we do not observe order-of-magnitude changes in the magnetic properties of either system 
demonstrates conclusively that the magnetic properties of the pristine materials are intrinsically 
2D, being determined almost exclusively by intralayer physics. The changes we do observe are a 
consequence of modulated intralayer magnetic interactions. In FePS$_3$, the drop in magnetic 
ordering temperature suggests a change in the competing FM and AFM interactions while the system 
remains in the Ising limit; the effects in NiPS$_3$ are far more subtle, with the interactions 
changing from weakly XY-type to almost isotropic (Heisenberg-type). We propose the charge doping that 
accompanies intercalation, which appears to be localized around the P ions and reaches a level of 
0.3-0.4~e$^-$/f.u., as the mechanism most likely to drive these changes. Our study of chemical 
intercalation with ionic liquids paves the way not only for approaching the 2D limit in van der Waals 
magnetic materials but also for tuning their intralayer magnetic interactions.

\begin{acknowledgments}
This work was supported by the National Key R\&D Program of China (under Grant Nos.~2023YFA1406500, 
2022YFA1402700, 2023YFA1406100, and 2019YFA0308602), the National Natural Science Foundation of China 
(under Grant Nos.~12134020, 12374156, 12174441, and 12074425), the Fundamental Research Funds for the 
Central Universities, and the Research Funds of Renmin University of China (under Grant Nos.~22XNH096 
and 23XNKJ22). Y.~F.~Guo acknowledges the Double First-Class Initiative Fund of ShanghaiTech University.
\end{acknowledgments}


\begin{thebibliography}{55}%
	\makeatletter
	\providecommand \@ifxundefined [1]{%
		\@ifx{#1\undefined}
	}%
	\providecommand \@ifnum [1]{%
		\ifnum #1\expandafter \@firstoftwo
		\else \expandafter \@secondoftwo
		\fi
	}%
	\providecommand \@ifx [1]{%
		\ifx #1\expandafter \@firstoftwo
		\else \expandafter \@secondoftwo
		\fi
	}%
	\providecommand \natexlab [1]{#1}%
	\providecommand \enquote  [1]{``#1''}%
	\providecommand \bibnamefont  [1]{#1}%
	\providecommand \bibfnamefont [1]{#1}%
	\providecommand \citenamefont [1]{#1}%
	\providecommand \href@noop [0]{\@secondoftwo}%
	\providecommand \href [0]{\begingroup \@sanitize@url \@href}%
	\providecommand \@href[1]{\@@startlink{#1}\@@href}%
	\providecommand \@@href[1]{\endgroup#1\@@endlink}%
	\providecommand \@sanitize@url [0]{\catcode `\\12\catcode `\$12\catcode
		`\&12\catcode `\#12\catcode `\^12\catcode `\_12\catcode `\%12\relax}%
	\providecommand \@@startlink[1]{}%
	\providecommand \@@endlink[0]{}%
	\providecommand \url  [0]{\begingroup\@sanitize@url \@url }%
	\providecommand \@url [1]{\endgroup\@href {#1}{\urlprefix }}%
	\providecommand \urlprefix  [0]{URL }%
	\providecommand \Eprint [0]{\href }%
	\providecommand \doibase [0]{http://dx.doi.org/}%
	\providecommand \selectlanguage [0]{\@gobble}%
	\providecommand \bibinfo  [0]{\@secondoftwo}%
	\providecommand \bibfield  [0]{\@secondoftwo}%
	\providecommand \translation [1]{[#1]}%
	\providecommand \BibitemOpen [0]{}%
	\providecommand \bibitemStop [0]{}%
	\providecommand \bibitemNoStop [0]{.\EOS\space}%
	\providecommand \EOS [0]{\spacefactor3000\relax}%
	\providecommand \BibitemShut  [1]{\csname bibitem#1\endcsname}%
	\let\auto@bib@innerbib\@empty
	\bibitem [{\citenamefont {Bethe}(1931)}]{Bethe_EPJA_1931}%
	\BibitemOpen
	\bibfield  {author} {\bibinfo {author} {\bibfnamefont {H.}~\bibnamefont
			{Bethe}},\ }\bibfield  {title} {\enquote {\bibinfo {title} {On the theory of
				metals.}}\ }\href@noop {} {\bibfield  {journal} {\bibinfo  {journal} {Eur. J.
				Phys. A}\ }\textbf {\bibinfo {volume} {71}},\ \bibinfo {pages} {205}
		(\bibinfo {year} {1931})}\BibitemShut {NoStop}%
	\bibitem [{\citenamefont {Alias}\ and\ \citenamefont
		{Sukumaran}(1992)}]{Joy_JACS_1992}%
	\BibitemOpen
	\bibfield  {author} {\bibinfo {author} {\bibfnamefont {J.~P.}\ \bibnamefont
			{Alias}}\ and\ \bibinfo {author} {\bibfnamefont {V.}~\bibnamefont
			{Sukumaran}},\ }\bibfield  {title} {\enquote {\bibinfo {title} {The
				intercalation reaction of pyridine with manganese thiophosphate,
				$\mathrm{MnPS}_3$},}\ }\href {\doibase 10.1021/ja00046a027} {\bibfield
		{journal} {\bibinfo  {journal} {J. Am. Chem. Soc}\ }\textbf {\bibinfo
			{volume} {114}},\ \bibinfo {pages} {7792} (\bibinfo {year}
		{1992})}\BibitemShut {NoStop}%
	\bibitem [{\citenamefont {Pattayil}\ and\ \citenamefont
		{Sukumaran}(1993)}]{Joy_ChemMater_1993}%
	\BibitemOpen
	\bibfield  {author} {\bibinfo {author} {\bibfnamefont {A.~J.}\ \bibnamefont
			{Pattayil}}\ and\ \bibinfo {author} {\bibfnamefont {V.}~\bibnamefont
			{Sukumaran}},\ }\bibfield  {title} {\enquote {\bibinfo {title}
			{{Intercalation of n-alkylamines in iron thiohypophosphate (FePS$_3$)}},}\
	}\href {\doibase 10.1021/cm00032a024} {\bibfield  {journal} {\bibinfo
			{journal} {Chem. Mater.}\ }\textbf {\bibinfo {volume} {5}},\ \bibinfo {pages}
		{1182} (\bibinfo {year} {1993})}\BibitemShut {NoStop}%
	\bibitem [{\citenamefont {Soumyanarayanan}\ \emph {et~al.}(2016)\citenamefont
		{Soumyanarayanan}, \citenamefont {Reyren}, \citenamefont {Fert},\ and\
		\citenamefont {Panagopoulos}}]{srfp_2016}%
	\BibitemOpen
	\bibfield  {author} {\bibinfo {author} {\bibfnamefont {A.}~\bibnamefont
			{Soumyanarayanan}}, \bibinfo {author} {\bibfnamefont {N.}~\bibnamefont
			{Reyren}}, \bibinfo {author} {\bibfnamefont {A.}~\bibnamefont {Fert}}, \ and\
		\bibinfo {author} {\bibfnamefont {C.}~\bibnamefont {Panagopoulos}},\
	}\bibfield  {title} {\enquote {\bibinfo {title} {{Emergent phenomena induced
					by spin-orbit coupling at surfaces and interfaces}},}\ }\href {\doibase
		10.1038/nature19820} {\bibfield  {journal} {\bibinfo  {journal} {Nature}\
		}\textbf {\bibinfo {volume} {539}},\ \bibinfo {pages} {509} (\bibinfo {year}
		{2016})}\BibitemShut {NoStop}%
	\bibitem [{\citenamefont {Burch}\ \emph {et~al.}(2018)\citenamefont {Burch},
		\citenamefont {Mandrus},\ and\ \citenamefont {Park}}]{burch_nature_2018}%
	\BibitemOpen
	\bibfield  {author} {\bibinfo {author} {\bibfnamefont {K.~S.}\ \bibnamefont
			{Burch}}, \bibinfo {author} {\bibfnamefont {D.}~\bibnamefont {Mandrus}}, \
		and\ \bibinfo {author} {\bibfnamefont {J.-G.}\ \bibnamefont {Park}},\
	}\bibfield  {title} {\enquote {\bibinfo {title} {{Magnetism in
					two-dimensional van der Waals materials}},}\ }\href {\doibase
		10.1038/s41586-018-0631-z} {\bibfield  {journal} {\bibinfo  {journal}
			{Nature}\ }\textbf {\bibinfo {volume} {563}},\ \bibinfo {pages} {47}
		(\bibinfo {year} {2018})}\BibitemShut {NoStop}%
	\bibitem [{\citenamefont {Gong}\ \emph {et~al.}(2017)\citenamefont {Gong},
		\citenamefont {Li}, \citenamefont {Li}, \citenamefont {Ji}, \citenamefont
		{Stern}, \citenamefont {Xia}, \citenamefont {Cao}, \citenamefont {Bao},
		\citenamefont {Wang}, \citenamefont {Wang}, \citenamefont {Qiu},
		\citenamefont {Cava}, \citenamefont {Louie}, \citenamefont {Xia},\ and\
		\citenamefont {Zhang}}]{Gong_Nature_2017}%
	\BibitemOpen
	\bibfield  {author} {\bibinfo {author} {\bibfnamefont {C.}~\bibnamefont
			{Gong}}, \bibinfo {author} {\bibfnamefont {L.}~\bibnamefont {Li}}, \bibinfo
		{author} {\bibfnamefont {Z.~L.}\ \bibnamefont {Li}}, \bibinfo {author}
		{\bibfnamefont {H.~W.}\ \bibnamefont {Ji}}, \bibinfo {author} {\bibfnamefont
			{A.}~\bibnamefont {Stern}}, \bibinfo {author} {\bibfnamefont
			{Y.}~\bibnamefont {Xia}}, \bibinfo {author} {\bibfnamefont {T.}~\bibnamefont
			{Cao}}, \bibinfo {author} {\bibfnamefont {W.}~\bibnamefont {Bao}}, \bibinfo
		{author} {\bibfnamefont {C.~Z.}\ \bibnamefont {Wang}}, \bibinfo {author}
		{\bibfnamefont {Y.}~\bibnamefont {Wang}}, \bibinfo {author} {\bibfnamefont
			{Z.~Q.}\ \bibnamefont {Qiu}}, \bibinfo {author} {\bibfnamefont {R.~J.}\
			\bibnamefont {Cava}}, \bibinfo {author} {\bibfnamefont {S.~G.}\ \bibnamefont
			{Louie}}, \bibinfo {author} {\bibfnamefont {J.}~\bibnamefont {Xia}}, \ and\
		\bibinfo {author} {\bibfnamefont {X.}~\bibnamefont {Zhang}},\ }\bibfield
	{title} {\enquote {\bibinfo {title} {{Discovery of intrinsic ferromagnetism
					in two-dimensional van der Waals crystals}},}\ }\href {\doibase
		10.1038/nature22060} {\bibfield  {journal} {\bibinfo  {journal} {Nature}\
		}\textbf {\bibinfo {volume} {546}},\ \bibinfo {pages} {265} (\bibinfo {year}
		{2017})}\BibitemShut {NoStop}%
	\bibitem [{\citenamefont {Huang}\ \emph {et~al.}(2017)\citenamefont {Huang},
		\citenamefont {Clark}, \citenamefont {Navarro-Moratalla}, \citenamefont
		{Klein}, \citenamefont {Cheng}, \citenamefont {Seyler}, \citenamefont
		{Zhong}, \citenamefont {Schmidgall}, \citenamefont {McGuire}, \citenamefont
		{Cobden}, \citenamefont {Yao}, \citenamefont {Xiao}, \citenamefont
		{Jarillo-Herrero},\ and\ \citenamefont {Xu}}]{Huang_Nature_2017}%
	\BibitemOpen
	\bibfield  {author} {\bibinfo {author} {\bibfnamefont {B.}~\bibnamefont
			{Huang}}, \bibinfo {author} {\bibfnamefont {G.}~\bibnamefont {Clark}},
		\bibinfo {author} {\bibfnamefont {E.}~\bibnamefont {Navarro-Moratalla}},
		\bibinfo {author} {\bibfnamefont {D.~R.}\ \bibnamefont {Klein}}, \bibinfo
		{author} {\bibfnamefont {R.}~\bibnamefont {Cheng}}, \bibinfo {author}
		{\bibfnamefont {K.~L.}\ \bibnamefont {Seyler}}, \bibinfo {author}
		{\bibfnamefont {D.}~\bibnamefont {Zhong}}, \bibinfo {author} {\bibfnamefont
			{E.}~\bibnamefont {Schmidgall}}, \bibinfo {author} {\bibfnamefont {M.~A.}\
			\bibnamefont {McGuire}}, \bibinfo {author} {\bibfnamefont {D.~H.}\
			\bibnamefont {Cobden}}, \bibinfo {author} {\bibfnamefont {W.}~\bibnamefont
			{Yao}}, \bibinfo {author} {\bibfnamefont {D.}~\bibnamefont {Xiao}}, \bibinfo
		{author} {\bibfnamefont {P.}~\bibnamefont {Jarillo-Herrero}}, \ and\ \bibinfo
		{author} {\bibfnamefont {X.~D.}\ \bibnamefont {Xu}},\ }\bibfield  {title}
	{\enquote {\bibinfo {title} {{Layer-dependent ferromagnetism in a van der
					Waals crystal down to the monolayer limit}},}\ }\href {\doibase
		10.1038/nature22391} {\bibfield  {journal} {\bibinfo  {journal} {Nature}\
		}\textbf {\bibinfo {volume} {546}},\ \bibinfo {pages} {270} (\bibinfo {year}
		{2017})}\BibitemShut {NoStop}%
	\bibitem [{\citenamefont {Tian}\ \emph {et~al.}(2016)\citenamefont {Tian},
		\citenamefont {Gray}, \citenamefont {Ji}, \citenamefont {Cava},\ and\
		\citenamefont {Burch}}]{Tian_2D_2016}%
	\BibitemOpen
	\bibfield  {author} {\bibinfo {author} {\bibfnamefont {Y.}~\bibnamefont
			{Tian}}, \bibinfo {author} {\bibfnamefont {M.~J.}\ \bibnamefont {Gray}},
		\bibinfo {author} {\bibfnamefont {H.~W.}\ \bibnamefont {Ji}}, \bibinfo
		{author} {\bibfnamefont {R.~J.}\ \bibnamefont {Cava}}, \ and\ \bibinfo
		{author} {\bibfnamefont {K.~S.}\ \bibnamefont {Burch}},\ }\bibfield  {title}
	{\enquote {\bibinfo {title} {{Magneto-elastic coupling in a potential
					ferromagnetic 2D atomic crystal}},}\ }\href {\doibase
		10.1088/2053-1583/3/2/025035} {\bibfield  {journal} {\bibinfo  {journal} {2D
				Mater.}\ }\textbf {\bibinfo {volume} {3}},\ \bibinfo {pages} {025035}
		(\bibinfo {year} {2016})}\BibitemShut {NoStop}%
	\bibitem [{\citenamefont {O'Hara}\ \emph {et~al.}(2018)\citenamefont {O'Hara},
		\citenamefont {Zhu}, \citenamefont {Trout}, \citenamefont {Ahmed},
		\citenamefont {Luo}, \citenamefont {Lee}, \citenamefont {Brenner},
		\citenamefont {Rajan}, \citenamefont {Gupta}, \citenamefont {McComb},\ and\
		\citenamefont {Kawakami}}]{Hara_2018_nanoLett}%
	\BibitemOpen
	\bibfield  {author} {\bibinfo {author} {\bibfnamefont {D.~J.}\ \bibnamefont
			{O'Hara}}, \bibinfo {author} {\bibfnamefont {T.~C.}\ \bibnamefont {Zhu}},
		\bibinfo {author} {\bibfnamefont {A.~H.}\ \bibnamefont {Trout}}, \bibinfo
		{author} {\bibfnamefont {A.~S.}\ \bibnamefont {Ahmed}}, \bibinfo {author}
		{\bibfnamefont {Y.~Q.~K.}\ \bibnamefont {Luo}}, \bibinfo {author}
		{\bibfnamefont {C.~H.}\ \bibnamefont {Lee}}, \bibinfo {author} {\bibfnamefont
			{M.~R.}\ \bibnamefont {Brenner}}, \bibinfo {author} {\bibfnamefont
			{S.}~\bibnamefont {Rajan}}, \bibinfo {author} {\bibfnamefont {J.~A.}\
			\bibnamefont {Gupta}}, \bibinfo {author} {\bibfnamefont {D.~W.}\ \bibnamefont
			{McComb}}, \ and\ \bibinfo {author} {\bibfnamefont {R.~K.}\ \bibnamefont
			{Kawakami}},\ }\bibfield  {title} {\enquote {\bibinfo {title} {{Room
					Temperature Intrinsic Ferromagnetism in Epitaxial Manganese Selenide Films in
					the Monolayer Limit}},}\ }\href {\doibase 10.1021/acs.nanolett.8b00683}
	{\bibfield  {journal} {\bibinfo  {journal} {Nano Lett.}\ }\textbf {\bibinfo
			{volume} {18}},\ \bibinfo {pages} {3125} (\bibinfo {year}
		{2018})}\BibitemShut {NoStop}%
	\bibitem [{\citenamefont {Bonilla}\ \emph {et~al.}(2018)\citenamefont
		{Bonilla}, \citenamefont {Kolekar}, \citenamefont {Ma}, \citenamefont {Diaz},
		\citenamefont {Kalappattil}, \citenamefont {Das}, \citenamefont {Eggers},
		\citenamefont {Gutierrez}, \citenamefont {Phan},\ and\ \citenamefont
		{Batzill}}]{Boni_NatNon_2018}%
	\BibitemOpen
	\bibfield  {author} {\bibinfo {author} {\bibfnamefont {M.}~\bibnamefont
			{Bonilla}}, \bibinfo {author} {\bibfnamefont {S.}~\bibnamefont {Kolekar}},
		\bibinfo {author} {\bibfnamefont {Y.~J.}\ \bibnamefont {Ma}}, \bibinfo
		{author} {\bibfnamefont {H.~C.}\ \bibnamefont {Diaz}}, \bibinfo {author}
		{\bibfnamefont {V.}~\bibnamefont {Kalappattil}}, \bibinfo {author}
		{\bibfnamefont {R.}~\bibnamefont {Das}}, \bibinfo {author} {\bibfnamefont
			{T.}~\bibnamefont {Eggers}}, \bibinfo {author} {\bibfnamefont {H.~R.}\
			\bibnamefont {Gutierrez}}, \bibinfo {author} {\bibfnamefont {M.-H.}\
			\bibnamefont {Phan}}, \ and\ \bibinfo {author} {\bibfnamefont
			{M.}~\bibnamefont {Batzill}},\ }\bibfield  {title} {\enquote {\bibinfo
			{title} {{Strong room-temperature ferromagnetism in VSe$_2$ monolayers on van
					der Waals substrates}},}\ }\href {\doibase 10.1038/s41565-018-0063-9}
	{\bibfield  {journal} {\bibinfo  {journal} {Nat. Nanotechnol.}\ }\textbf
		{\bibinfo {volume} {13}},\ \bibinfo {pages} {289} (\bibinfo {year}
		{2018})}\BibitemShut {NoStop}%
	\bibitem [{\citenamefont {Le~Flem}\ \emph {et~al.}(1982)\citenamefont
		{Le~Flem}, \citenamefont {Brec}, \citenamefont {Ouvard}, \citenamefont
		{Louisy},\ and\ \citenamefont {Segransan}}]{Flem_JPCS_1982}%
	\BibitemOpen
	\bibfield  {author} {\bibinfo {author} {\bibfnamefont {G.}~\bibnamefont
			{Le~Flem}}, \bibinfo {author} {\bibfnamefont {R.}~\bibnamefont {Brec}},
		\bibinfo {author} {\bibfnamefont {G.}~\bibnamefont {Ouvard}}, \bibinfo
		{author} {\bibfnamefont {A.}~\bibnamefont {Louisy}}, \ and\ \bibinfo {author}
		{\bibfnamefont {P.}~\bibnamefont {Segransan}},\ }\bibfield  {title} {\enquote
		{\bibinfo {title} {{Magnetic interactions in the layer compounds MPX$_3$ (M =
					Mn, Fe, and Ni; X = S, Se)}},}\ }\href {\doibase
		https://doi.org/10.1016/0022-3697(82)90156-1} {\bibfield  {journal} {\bibinfo
			{journal} {J. Phys. Chem. Solids}\ }\textbf {\bibinfo {volume} {43}},\
		\bibinfo {pages} {455} (\bibinfo {year} {1982})}\BibitemShut {NoStop}%
	\bibitem [{\citenamefont {Ouvrard}\ \emph {et~al.}(1985)\citenamefont
		{Ouvrard}, \citenamefont {Brec},\ and\ \citenamefont
		{Rouxel}}]{Ouvrard_MRB_1985}%
	\BibitemOpen
	\bibfield  {author} {\bibinfo {author} {\bibfnamefont {G.}~\bibnamefont
			{Ouvrard}}, \bibinfo {author} {\bibfnamefont {R.}~\bibnamefont {Brec}}, \
		and\ \bibinfo {author} {\bibfnamefont {J.}~\bibnamefont {Rouxel}},\
	}\bibfield  {title} {\enquote {\bibinfo {title} {{Structural determination of
					some MPS$_3$ layered phases (M = Mn, Fe, Co, Ni, and Cd)}},}\ }\href
	{\doibase https://doi.org/10.1016/0025-5408(85)90092-3} {\bibfield  {journal}
		{\bibinfo  {journal} {Mater. Res. Bull.}\ }\textbf {\bibinfo {volume} {20}},\
		\bibinfo {pages} {1181} (\bibinfo {year} {1985})}\BibitemShut {NoStop}%
	\bibitem [{\citenamefont {Joy}\ and\ \citenamefont
		{Vasudevan}(1992)}]{Joy_PRB_1992}%
	\BibitemOpen
	\bibfield  {author} {\bibinfo {author} {\bibfnamefont {P.~A.}\ \bibnamefont
			{Joy}}\ and\ \bibinfo {author} {\bibfnamefont {S.}~\bibnamefont
			{Vasudevan}},\ }\bibfield  {title} {\enquote {\bibinfo {title} {{Magnetism in
					the layered transition-metal thiophosphates MPS$_3$ (M = Mn, Fe, and Ni)}},}\
	}\href {\doibase 10.1103/PhysRevB.46.5425} {\bibfield  {journal} {\bibinfo
			{journal} {Phys. Rev. B}\ }\textbf {\bibinfo {volume} {46}},\ \bibinfo
		{pages} {5425} (\bibinfo {year} {1992})}\BibitemShut {NoStop}%
	\bibitem [{\citenamefont {Chittari}\ \emph {et~al.}(2016)\citenamefont
		{Chittari}, \citenamefont {Park}, \citenamefont {Lee}, \citenamefont {Han},
		\citenamefont {MacDonald}, \citenamefont {Hwang},\ and\ \citenamefont
		{Jung}}]{Chittari_PRB_2016}%
	\BibitemOpen
	\bibfield  {author} {\bibinfo {author} {\bibfnamefont {B.~L.}\ \bibnamefont
			{Chittari}}, \bibinfo {author} {\bibfnamefont {Y.}~\bibnamefont {Park}},
		\bibinfo {author} {\bibfnamefont {D.}~\bibnamefont {Lee}}, \bibinfo {author}
		{\bibfnamefont {M.}~\bibnamefont {Han}}, \bibinfo {author} {\bibfnamefont
			{A.~H.}\ \bibnamefont {MacDonald}}, \bibinfo {author} {\bibfnamefont
			{E.}~\bibnamefont {Hwang}}, \ and\ \bibinfo {author} {\bibfnamefont
			{J.}~\bibnamefont {Jung}},\ }\bibfield  {title} {\enquote {\bibinfo {title}
			{Electronic and magnetic properties of single-layer $\mathrm{MPX}_{3}$ metal
				phosphorous trichalcogenides},}\ }\href {\doibase 10.1103/PhysRevB.94.184428}
	{\bibfield  {journal} {\bibinfo  {journal} {Phys. Rev. B}\ }\textbf {\bibinfo
			{volume} {94}},\ \bibinfo {pages} {184428} (\bibinfo {year}
		{2016})}\BibitemShut {NoStop}%
	\bibitem [{\citenamefont {Du}\ \emph {et~al.}(2016)\citenamefont {Du},
		\citenamefont {Wang}, \citenamefont {Liu}, \citenamefont {Hu}, \citenamefont
		{Utama}, \citenamefont {Gan}, \citenamefont {Xiong},\ and\ \citenamefont
		{Kloc}}]{Chris_ACSNano_2016}%
	\BibitemOpen
	\bibfield  {author} {\bibinfo {author} {\bibfnamefont {K.~Z.}\ \bibnamefont
			{Du}}, \bibinfo {author} {\bibfnamefont {X.~Z.}\ \bibnamefont {Wang}},
		\bibinfo {author} {\bibfnamefont {Y.}~\bibnamefont {Liu}}, \bibinfo {author}
		{\bibfnamefont {P.}~\bibnamefont {Hu}}, \bibinfo {author} {\bibfnamefont
			{M.~I.~B.}\ \bibnamefont {Utama}}, \bibinfo {author} {\bibfnamefont {C.~K.}\
			\bibnamefont {Gan}}, \bibinfo {author} {\bibfnamefont {Q.~H.}\ \bibnamefont
			{Xiong}}, \ and\ \bibinfo {author} {\bibfnamefont {C.}~\bibnamefont {Kloc}},\
	}\bibfield  {title} {\enquote {\bibinfo {title} {{Weak van der Waals
					Stacking, Wide-Range Band Gap, and Raman Study on Ultrathin Layers of Metal
					Phosphorus Trichalcogenides}},}\ }\href {\doibase 10.1021/acsnano.5b05927}
	{\bibfield  {journal} {\bibinfo  {journal} {ACS Nano}\ }\textbf {\bibinfo
			{volume} {10}},\ \bibinfo {pages} {1738} (\bibinfo {year}
		{2016})}\BibitemShut {NoStop}%
	\bibitem [{\citenamefont {Wang}\ \emph {et~al.}(2018)\citenamefont {Wang},
		\citenamefont {Shifa}, \citenamefont {Yu}, \citenamefont {He}, \citenamefont
		{Liu}, \citenamefont {Wang}, \citenamefont {Wang}, \citenamefont {Zhan},
		\citenamefont {Lou}, \citenamefont {Xia},\ and\ \citenamefont
		{He}}]{Wang_AdvML_2018}%
	\BibitemOpen
	\bibfield  {author} {\bibinfo {author} {\bibfnamefont {F.~M.}\ \bibnamefont
			{Wang}}, \bibinfo {author} {\bibfnamefont {T.~A.}\ \bibnamefont {Shifa}},
		\bibinfo {author} {\bibfnamefont {P.}~\bibnamefont {Yu}}, \bibinfo {author}
		{\bibfnamefont {P.}~\bibnamefont {He}}, \bibinfo {author} {\bibfnamefont
			{Y.}~\bibnamefont {Liu}}, \bibinfo {author} {\bibfnamefont {F.}~\bibnamefont
			{Wang}}, \bibinfo {author} {\bibfnamefont {Z.~X.}\ \bibnamefont {Wang}},
		\bibinfo {author} {\bibfnamefont {X.~Y.}\ \bibnamefont {Zhan}}, \bibinfo
		{author} {\bibfnamefont {X.~D.}\ \bibnamefont {Lou}}, \bibinfo {author}
		{\bibfnamefont {F.}~\bibnamefont {Xia}}, \ and\ \bibinfo {author}
		{\bibfnamefont {J.}~\bibnamefont {He}},\ }\bibfield  {title} {\enquote
		{\bibinfo {title} {{New Frontiers on van der Waals Layered Metal Phosphorous
					Trichalcogenides}},}\ }\href {\doibase
		https://doi.org/10.1002/adfm.201802151} {\bibfield  {journal} {\bibinfo
			{journal} {Adv. Funct. Mater.}\ }\textbf {\bibinfo {volume} {28}},\ \bibinfo
		{pages} {1802151} (\bibinfo {year} {2018})}\BibitemShut {NoStop}%
	\bibitem [{\citenamefont {Haines}\ \emph {et~al.}(2018)\citenamefont {Haines},
		\citenamefont {Coak}, \citenamefont {Wildes}, \citenamefont {Lampronti},
		\citenamefont {Liu}, \citenamefont {Nahai-Williamson}, \citenamefont
		{Hamidov}, \citenamefont {Daisenberger},\ and\ \citenamefont
		{Saxena}}]{Haines_PRL_2018}%
	\BibitemOpen
	\bibfield  {author} {\bibinfo {author} {\bibfnamefont {C.~R.~S.}\
			\bibnamefont {Haines}}, \bibinfo {author} {\bibfnamefont {M.~J.}\
			\bibnamefont {Coak}}, \bibinfo {author} {\bibfnamefont {A.~R.}\ \bibnamefont
			{Wildes}}, \bibinfo {author} {\bibfnamefont {G.~I.}\ \bibnamefont
			{Lampronti}}, \bibinfo {author} {\bibfnamefont {C.}~\bibnamefont {Liu}},
		\bibinfo {author} {\bibfnamefont {P.}~\bibnamefont {Nahai-Williamson}},
		\bibinfo {author} {\bibfnamefont {H.}~\bibnamefont {Hamidov}}, \bibinfo
		{author} {\bibfnamefont {D.}~\bibnamefont {Daisenberger}}, \ and\ \bibinfo
		{author} {\bibfnamefont {S.~S.}\ \bibnamefont {Saxena}},\ }\bibfield  {title}
	{\enquote {\bibinfo {title} {{Pressure-Induced Electronic and Structural
					Phase Evolution in the van der Waals Compound $\mathrm{FePS}_{3}$}},}\ }\href
	{\doibase 10.1103/PhysRevLett.121.266801} {\bibfield  {journal} {\bibinfo
			{journal} {Phys. Rev. Lett.}\ }\textbf {\bibinfo {volume} {121}},\ \bibinfo
		{pages} {266801} (\bibinfo {year} {2018})}\BibitemShut {NoStop}%
	\bibitem [{\citenamefont {Kang}\ \emph {et~al.}(2020)\citenamefont {Kang},
		\citenamefont {Kim}, \citenamefont {Kim}, \citenamefont {Kim}, \citenamefont
		{Sim}, \citenamefont {Lee}, \citenamefont {Lee}, \citenamefont {Park},
		\citenamefont {Yun}, \citenamefont {Kim}, \citenamefont {Nag}, \citenamefont
		{Walters}, \citenamefont {Garcia-Fernandez}, \citenamefont {Li},
		\citenamefont {Chapon}, \citenamefont {Zhou}, \citenamefont {Son},
		\citenamefont {Kim}, \citenamefont {Cheong},\ and\ \citenamefont
		{Park}}]{kang_nature_2020}%
	\BibitemOpen
	\bibfield  {author} {\bibinfo {author} {\bibfnamefont {S.}~\bibnamefont
			{Kang}}, \bibinfo {author} {\bibfnamefont {K.}~\bibnamefont {Kim}}, \bibinfo
		{author} {\bibfnamefont {B.~H.}\ \bibnamefont {Kim}}, \bibinfo {author}
		{\bibfnamefont {J.}~\bibnamefont {Kim}}, \bibinfo {author} {\bibfnamefont
			{K.~I.}\ \bibnamefont {Sim}}, \bibinfo {author} {\bibfnamefont {J.-U.}\
			\bibnamefont {Lee}}, \bibinfo {author} {\bibfnamefont {S.}~\bibnamefont
			{Lee}}, \bibinfo {author} {\bibfnamefont {K.}~\bibnamefont {Park}}, \bibinfo
		{author} {\bibfnamefont {S.}~\bibnamefont {Yun}}, \bibinfo {author}
		{\bibfnamefont {T.}~\bibnamefont {Kim}}, \bibinfo {author} {\bibfnamefont
			{A.}~\bibnamefont {Nag}}, \bibinfo {author} {\bibfnamefont {A.}~\bibnamefont
			{Walters}}, \bibinfo {author} {\bibfnamefont {M.}~\bibnamefont
			{Garcia-Fernandez}}, \bibinfo {author} {\bibfnamefont {J.}~\bibnamefont
			{Li}}, \bibinfo {author} {\bibfnamefont {L.}~\bibnamefont {Chapon}}, \bibinfo
		{author} {\bibfnamefont {K.-J.}\ \bibnamefont {Zhou}}, \bibinfo {author}
		{\bibfnamefont {Y.-W.}\ \bibnamefont {Son}}, \bibinfo {author} {\bibfnamefont
			{J.~H.}\ \bibnamefont {Kim}}, \bibinfo {author} {\bibfnamefont
			{H.}~\bibnamefont {Cheong}}, \ and\ \bibinfo {author} {\bibfnamefont {J.-G.}\
			\bibnamefont {Park}},\ }\bibfield  {title} {\enquote {\bibinfo {title}
			{{Coherent many-body exciton in van der Waals antiferromagnet NiPS$_3$}},}\
	}\href {\doibase 10.1038/s41586-020-2520-5} {\bibfield  {journal} {\bibinfo
			{journal} {Nature}\ }\textbf {\bibinfo {volume} {583}},\ \bibinfo {pages}
		{785} (\bibinfo {year} {2020})}\BibitemShut {NoStop}%
	\bibitem [{\citenamefont {Kurosawa}\ \emph {et~al.}(1983)\citenamefont
		{Kurosawa}, \citenamefont {Saito},\ and\ \citenamefont {Yasuo}}]{Yasuo_1983}%
	\BibitemOpen
	\bibfield  {author} {\bibinfo {author} {\bibfnamefont {K.}~\bibnamefont
			{Kurosawa}}, \bibinfo {author} {\bibfnamefont {S.}~\bibnamefont {Saito}}, \
		and\ \bibinfo {author} {\bibfnamefont {Y.}~\bibnamefont {Yasuo}},\ }\bibfield
	{title} {\enquote {\bibinfo {title} {Neutron diffraction study on
				$\mathrm{MnPS}_3$ and $\mathrm{FePS}_3$},}\ }\href {\doibase
		10.1143/JPSJ.52.3919} {\bibfield  {journal} {\bibinfo  {journal} {J. Phys.
				Soc. Jpn.}\ }\textbf {\bibinfo {volume} {52}},\ \bibinfo {pages} {3919}
		(\bibinfo {year} {1983})}\BibitemShut {NoStop}%
	\bibitem [{\citenamefont {Chandrasekharan}\ and\ \citenamefont
		{Vasudevan}(1994)}]{Pramana_1994}%
	\BibitemOpen
	\bibfield  {author} {\bibinfo {author} {\bibfnamefont {N.}~\bibnamefont
			{Chandrasekharan}}\ and\ \bibinfo {author} {\bibfnamefont {S.}~\bibnamefont
			{Vasudevan}},\ }\bibfield  {title} {\enquote {\bibinfo {title} {Magnetism,
				exchange and crystal field parameters in the orbitally unquenched
				$\mathrm{I}$sing antiferromagnet $\mathrm{Fe}\mathrm{PS}_3$},}\ }\href
	{\doibase https://doi.org/10.1007/BF02847596} {\bibfield  {journal} {\bibinfo
			{journal} {Pramana}\ }\textbf {\bibinfo {volume} {43}},\ \bibinfo {pages}
		{21} (\bibinfo {year} {1994})}\BibitemShut {NoStop}%
	\bibitem [{\citenamefont {Chatterjee}(1995)}]{Chatterjee_PRB_1995}%
	\BibitemOpen
	\bibfield  {author} {\bibinfo {author} {\bibfnamefont {I.}~\bibnamefont
			{Chatterjee}},\ }\bibfield  {title} {\enquote {\bibinfo {title} {Magnetic
				properties of layered antiferromagnets},}\ }\href {\doibase
		10.1103/PhysRevB.51.3937} {\bibfield  {journal} {\bibinfo  {journal} {Phys.
				Rev. B}\ }\textbf {\bibinfo {volume} {51}},\ \bibinfo {pages} {3937}
		(\bibinfo {year} {1995})}\BibitemShut {NoStop}%
	\bibitem [{\citenamefont {Rule}\ \emph {et~al.}(2007)\citenamefont {Rule},
		\citenamefont {McIntyre}, \citenamefont {Kennedy},\ and\ \citenamefont
		{Hicks}}]{Hicks_PRB_2007}%
	\BibitemOpen
	\bibfield  {author} {\bibinfo {author} {\bibfnamefont {K.~C.}\ \bibnamefont
			{Rule}}, \bibinfo {author} {\bibfnamefont {G.~J.}\ \bibnamefont {McIntyre}},
		\bibinfo {author} {\bibfnamefont {S.~J.}\ \bibnamefont {Kennedy}}, \ and\
		\bibinfo {author} {\bibfnamefont {T.~J.}\ \bibnamefont {Hicks}},\ }\bibfield
	{title} {\enquote {\bibinfo {title} {Single-crystal and powder neutron
				diffraction experiments on $\mathrm{Fe}\mathrm{PS}_{3}$: Search for the
				magnetic structure},}\ }\href {\doibase 10.1103/PhysRevB.76.134402}
	{\bibfield  {journal} {\bibinfo  {journal} {Phys. Rev. B}\ }\textbf {\bibinfo
			{volume} {76}},\ \bibinfo {pages} {134402} (\bibinfo {year}
		{2007})}\BibitemShut {NoStop}%
	\bibitem [{\citenamefont {Lee}\ \emph {et~al.}(2016)\citenamefont {Lee},
		\citenamefont {Lee}, \citenamefont {Ryoo}, \citenamefont {Kang},
		\citenamefont {Kim}, \citenamefont {Kim}, \citenamefont {Park}, \citenamefont
		{Park},\ and\ \citenamefont {Cheong}}]{Lee_NL_2016}%
	\BibitemOpen
	\bibfield  {author} {\bibinfo {author} {\bibfnamefont {J.~U.}\ \bibnamefont
			{Lee}}, \bibinfo {author} {\bibfnamefont {S.}~\bibnamefont {Lee}}, \bibinfo
		{author} {\bibfnamefont {J.~H.}\ \bibnamefont {Ryoo}}, \bibinfo {author}
		{\bibfnamefont {S.}~\bibnamefont {Kang}}, \bibinfo {author} {\bibfnamefont
			{T.~Y.}\ \bibnamefont {Kim}}, \bibinfo {author} {\bibfnamefont
			{P.}~\bibnamefont {Kim}}, \bibinfo {author} {\bibfnamefont {C.~H.}\
			\bibnamefont {Park}}, \bibinfo {author} {\bibfnamefont {J.~G.}\ \bibnamefont
			{Park}}, \ and\ \bibinfo {author} {\bibfnamefont {H.}~\bibnamefont
			{Cheong}},\ }\bibfield  {title} {\enquote {\bibinfo {title} {{Ising-Type
					Magnetic Ordering in Atomically Thin FePS$_3$}},}\ }\href {\doibase
		10.1021/acs.nanolett.6b03052} {\bibfield  {journal} {\bibinfo  {journal}
			{Nano Lett.}\ }\textbf {\bibinfo {volume} {16}},\ \bibinfo {pages} {7433}
		(\bibinfo {year} {2016})}\BibitemShut {NoStop}%
	\bibitem [{\citenamefont {Wildes}\ \emph {et~al.}(2017)\citenamefont {Wildes},
		\citenamefont {Simonet}, \citenamefont {Ressouche}, \citenamefont {Ballou},\
		and\ \citenamefont {McIntyre}}]{Wildes_JPCM_2017}%
	\BibitemOpen
	\bibfield  {author} {\bibinfo {author} {\bibfnamefont {A.~R.}\ \bibnamefont
			{Wildes}}, \bibinfo {author} {\bibfnamefont {V.}~\bibnamefont {Simonet}},
		\bibinfo {author} {\bibfnamefont {E.}~\bibnamefont {Ressouche}}, \bibinfo
		{author} {\bibfnamefont {R.}~\bibnamefont {Ballou}}, \ and\ \bibinfo {author}
		{\bibfnamefont {G.~J.}\ \bibnamefont {McIntyre}},\ }\bibfield  {title}
	{\enquote {\bibinfo {title} {{The magnetic properties and structure of the
					quasi-two-dimensional antiferromagnet $\mathrm{CoPS}_3$}},}\ }\href {\doibase
		10.1088/1361-648X/aa8a43} {\bibfield  {journal} {\bibinfo  {journal} {J.
				Phys. Condens. Matter}\ }\textbf {\bibinfo {volume} {29}},\ \bibinfo {pages}
		{455801} (\bibinfo {year} {2017})}\BibitemShut {NoStop}%
	\bibitem [{\citenamefont {Brec}(1986)}]{BREC_SSI_1986}%
	\BibitemOpen
	\bibfield  {author} {\bibinfo {author} {\bibfnamefont {R.}~\bibnamefont
			{Brec}},\ }\bibfield  {title} {\enquote {\bibinfo {title} {{Review on
					structural and chemical properties of transition metal phosphorous
					trisulfides MPS$_3$}},}\ }\href {\doibase
		https://doi.org/10.1016/0167-2738(86)90055-X} {\bibfield  {journal} {\bibinfo
			{journal} {Solid State Ion.}\ }\textbf {\bibinfo {volume} {22}},\ \bibinfo
		{pages} {3} (\bibinfo {year} {1986})}\BibitemShut {NoStop}%
	\bibitem [{\citenamefont {Kuo}\ \emph {et~al.}(2016)\citenamefont {Kuo},
		\citenamefont {Neumann}, \citenamefont {Balamurugan}, \citenamefont {Park},
		\citenamefont {Kang}, \citenamefont {Shiu}, \citenamefont {Kang},
		\citenamefont {Hong}, \citenamefont {Han}, \citenamefont {Noh},\ and\
		\citenamefont {Park}}]{Kuo_SciRep_2016}%
	\BibitemOpen
	\bibfield  {author} {\bibinfo {author} {\bibfnamefont {C.~T.}\ \bibnamefont
			{Kuo}}, \bibinfo {author} {\bibfnamefont {M.}~\bibnamefont {Neumann}},
		\bibinfo {author} {\bibfnamefont {K.}~\bibnamefont {Balamurugan}}, \bibinfo
		{author} {\bibfnamefont {H.~J.}\ \bibnamefont {Park}}, \bibinfo {author}
		{\bibfnamefont {S.}~\bibnamefont {Kang}}, \bibinfo {author} {\bibfnamefont
			{H.~W.}\ \bibnamefont {Shiu}}, \bibinfo {author} {\bibfnamefont {J.~H.}\
			\bibnamefont {Kang}}, \bibinfo {author} {\bibfnamefont {B.~H.}\ \bibnamefont
			{Hong}}, \bibinfo {author} {\bibfnamefont {M.}~\bibnamefont {Han}}, \bibinfo
		{author} {\bibfnamefont {T.~W.}\ \bibnamefont {Noh}}, \ and\ \bibinfo
		{author} {\bibfnamefont {J.-G.}\ \bibnamefont {Park}},\ }\bibfield  {title}
	{\enquote {\bibinfo {title} {{Exfoliation and Raman Spectroscopic Fingerprint
					of Few-Layer $\mathrm{NiPS}_{3}$ van der Waals Crystals}},}\ }\href {\doibase
		10.1038/srep20904} {\bibfield  {journal} {\bibinfo  {journal} {Sci. Rep.}\
		}\textbf {\bibinfo {volume} {6}},\ \bibinfo {pages} {20904} (\bibinfo {year}
		{2016})}\BibitemShut {NoStop}%
	\bibitem [{\citenamefont {Wang}\ \emph {et~al.}(2016)\citenamefont {Wang},
		\citenamefont {Du}, \citenamefont {Liu}, \citenamefont {Hu}, \citenamefont
		{Zhang}, \citenamefont {Zhang}, \citenamefont {Owen}, \citenamefont {Lu},
		\citenamefont {Gan}, \citenamefont {Sengupta}, \citenamefont {Kloc},\ and\
		\citenamefont {Xiong}}]{Wang_2DMaterial_2016}%
	\BibitemOpen
	\bibfield  {author} {\bibinfo {author} {\bibfnamefont {X.~Z.}\ \bibnamefont
			{Wang}}, \bibinfo {author} {\bibfnamefont {K.~Z.}\ \bibnamefont {Du}},
		\bibinfo {author} {\bibfnamefont {Y.~Y.~F.}\ \bibnamefont {Liu}}, \bibinfo
		{author} {\bibfnamefont {P.}~\bibnamefont {Hu}}, \bibinfo {author}
		{\bibfnamefont {J.}~\bibnamefont {Zhang}}, \bibinfo {author} {\bibfnamefont
			{Q.}~\bibnamefont {Zhang}}, \bibinfo {author} {\bibfnamefont {M.~H.~S.}\
			\bibnamefont {Owen}}, \bibinfo {author} {\bibfnamefont {X.}~\bibnamefont
			{Lu}}, \bibinfo {author} {\bibfnamefont {C.W.}\ \bibnamefont {Gan}}, \bibinfo
		{author} {\bibfnamefont {P.}~\bibnamefont {Sengupta}}, \bibinfo {author}
		{\bibfnamefont {C.}~\bibnamefont {Kloc}}, \ and\ \bibinfo {author}
		{\bibfnamefont {Q.~H.}\ \bibnamefont {Xiong}},\ }\bibfield  {title} {\enquote
		{\bibinfo {title} {Raman spectroscopy of atomically thin two-dimensional
				magnetic iron phosphorus trisulfide ($\mathrm{FePS}_3$) crystals},}\ }\href
	{\doibase 10.1088/2053-1583/3/3/031009} {\bibfield  {journal} {\bibinfo
			{journal} {2D Mater.}\ }\textbf {\bibinfo {volume} {3}},\ \bibinfo {pages}
		{031009} (\bibinfo {year} {2016})}\BibitemShut {NoStop}%
	\bibitem [{\citenamefont {McCreary}\ \emph {et~al.}(2020)\citenamefont
		{McCreary}, \citenamefont {Simpson}, \citenamefont {Mai}, \citenamefont
		{McMichael}, \citenamefont {Douglas}, \citenamefont {Butch}, \citenamefont
		{Dennis}, \citenamefont {Vald\'es~Aguilar},\ and\ \citenamefont
		{Hight~Walker}}]{Walker_PRB_2020}%
	\BibitemOpen
	\bibfield  {author} {\bibinfo {author} {\bibfnamefont {A.}~\bibnamefont
			{McCreary}}, \bibinfo {author} {\bibfnamefont {J.~R.}\ \bibnamefont
			{Simpson}}, \bibinfo {author} {\bibfnamefont {T.~T.}\ \bibnamefont {Mai}},
		\bibinfo {author} {\bibfnamefont {R.~D.}\ \bibnamefont {McMichael}}, \bibinfo
		{author} {\bibfnamefont {J.~E.}\ \bibnamefont {Douglas}}, \bibinfo {author}
		{\bibfnamefont {N.}~\bibnamefont {Butch}}, \bibinfo {author} {\bibfnamefont
			{C.}~\bibnamefont {Dennis}}, \bibinfo {author} {\bibfnamefont
			{R.}~\bibnamefont {Vald\'es~Aguilar}}, \ and\ \bibinfo {author}
		{\bibfnamefont {A.~R.}\ \bibnamefont {Hight~Walker}},\ }\bibfield  {title}
	{\enquote {\bibinfo {title} {{Quasi-two-dimensional magnon identification in
					antiferromagnetic $\mathrm{FePS}_{3}$ via magneto-Raman spectroscopy}},}\
	}\href {\doibase 10.1103/PhysRevB.101.064416} {\bibfield  {journal} {\bibinfo
			{journal} {Phys. Rev. B}\ }\textbf {\bibinfo {volume} {101}},\ \bibinfo
		{pages} {064416} (\bibinfo {year} {2020})}\BibitemShut {NoStop}%
	\bibitem [{\citenamefont {Hu}\ \emph {et~al.}(2023)\citenamefont {Hu},
		\citenamefont {Wang}, \citenamefont {Chen}, \citenamefont {Xu}, \citenamefont
		{Li}, \citenamefont {Liu}, \citenamefont {Gu}, \citenamefont {Wang},
		\citenamefont {Zhang}, \citenamefont {Yao},\ and\ \citenamefont
		{Xiong}}]{Hu_PRB_2023}%
	\BibitemOpen
	\bibfield  {author} {\bibinfo {author} {\bibfnamefont {L.~L.}\ \bibnamefont
			{Hu}}, \bibinfo {author} {\bibfnamefont {H.~X.}\ \bibnamefont {Wang}},
		\bibinfo {author} {\bibfnamefont {Y.~Z.}\ \bibnamefont {Chen}}, \bibinfo
		{author} {\bibfnamefont {K.}~\bibnamefont {Xu}}, \bibinfo {author}
		{\bibfnamefont {M.-R.}\ \bibnamefont {Li}}, \bibinfo {author} {\bibfnamefont
			{H.~Y.}\ \bibnamefont {Liu}}, \bibinfo {author} {\bibfnamefont
			{P.}~\bibnamefont {Gu}}, \bibinfo {author} {\bibfnamefont {Y.~B.}\
			\bibnamefont {Wang}}, \bibinfo {author} {\bibfnamefont {M.~D.}\ \bibnamefont
			{Zhang}}, \bibinfo {author} {\bibfnamefont {H.}~\bibnamefont {Yao}}, \ and\
		\bibinfo {author} {\bibfnamefont {Q.~H.}\ \bibnamefont {Xiong}},\ }\bibfield
	{title} {\enquote {\bibinfo {title} {Observation of a magnetic phase
				transition in monolayer $\mathrm{NiPS}_3$},}\ }\href {\doibase
		10.1103/PhysRevB.107.L220407} {\bibfield  {journal} {\bibinfo  {journal}
			{Phys. Rev. B}\ }\textbf {\bibinfo {volume} {107}},\ \bibinfo {pages}
		{L220407} (\bibinfo {year} {2023})}\BibitemShut {NoStop}%
	\bibitem [{\citenamefont {Zhang}\ \emph {et~al.}(2020)\citenamefont {Zhang},
		\citenamefont {Rousuli}, \citenamefont {Shen}, \citenamefont {Zhang},
		\citenamefont {Wang}, \citenamefont {Luo}, \citenamefont {Wang},
		\citenamefont {Wu}, \citenamefont {Xu}, \citenamefont {Duan}, \citenamefont
		{Yao}, \citenamefont {Yu},\ and\ \citenamefont {Zhou}}]{Zhou_SB_2020}%
	\BibitemOpen
	\bibfield  {author} {\bibinfo {author} {\bibfnamefont {H.~X.}\ \bibnamefont
			{Zhang}}, \bibinfo {author} {\bibfnamefont {A.}~\bibnamefont {Rousuli}},
		\bibinfo {author} {\bibfnamefont {S.~C.}\ \bibnamefont {Shen}}, \bibinfo
		{author} {\bibfnamefont {K.~N.}\ \bibnamefont {Zhang}}, \bibinfo {author}
		{\bibfnamefont {C.}~\bibnamefont {Wang}}, \bibinfo {author} {\bibfnamefont
			{L.~P.}\ \bibnamefont {Luo}}, \bibinfo {author} {\bibfnamefont {J.~Z.}\
			\bibnamefont {Wang}}, \bibinfo {author} {\bibfnamefont {Y.}~\bibnamefont
			{Wu}}, \bibinfo {author} {\bibfnamefont {Y.}~\bibnamefont {Xu}}, \bibinfo
		{author} {\bibfnamefont {W.~H.}\ \bibnamefont {Duan}}, \bibinfo {author}
		{\bibfnamefont {H.}~\bibnamefont {Yao}}, \bibinfo {author} {\bibfnamefont
			{P.}~\bibnamefont {Yu}}, \ and\ \bibinfo {author} {\bibfnamefont {S.~Y.}\
			\bibnamefont {Zhou}},\ }\bibfield  {title} {\enquote {\bibinfo {title}
			{Enhancement of superconductivity in organic-inorganic hybrid topological
				materials},}\ }\href {\doibase https://doi.org/10.1016/j.scib.2019.11.021}
	{\bibfield  {journal} {\bibinfo  {journal} {Sci. Bull.}\ }\textbf {\bibinfo
			{volume} {65}},\ \bibinfo {pages} {188} (\bibinfo {year} {2020})}\BibitemShut
	{NoStop}%
	\bibitem [{\citenamefont {Zhang}\ \emph {et~al.}(2022)\citenamefont {Zhang},
		\citenamefont {Rousuli}, \citenamefont {Zhang}, \citenamefont {Luo},
		\citenamefont {Guo}, \citenamefont {Cong}, \citenamefont {Lin}, \citenamefont
		{Bao}, \citenamefont {Zhang}, \citenamefont {Xu}, \citenamefont {Feng},
		\citenamefont {Shen}, \citenamefont {Zhao}, \citenamefont {Yao},
		\citenamefont {Wu}, \citenamefont {Ji}, \citenamefont {Chen}, \citenamefont
		{Tan}, \citenamefont {Xue}, \citenamefont {Xu}, \citenamefont {Duan},
		\citenamefont {Yu},\ and\ \citenamefont {Zhou}}]{Zhou_NP_2022}%
	\BibitemOpen
	\bibfield  {author} {\bibinfo {author} {\bibfnamefont {H.~X.}\ \bibnamefont
			{Zhang}}, \bibinfo {author} {\bibfnamefont {A.}~\bibnamefont {Rousuli}},
		\bibinfo {author} {\bibfnamefont {K.~N.}\ \bibnamefont {Zhang}}, \bibinfo
		{author} {\bibfnamefont {L.~P.}\ \bibnamefont {Luo}}, \bibinfo {author}
		{\bibfnamefont {C.~G.}\ \bibnamefont {Guo}}, \bibinfo {author} {\bibfnamefont
			{X.}~\bibnamefont {Cong}}, \bibinfo {author} {\bibfnamefont {Z.~Z.}\
			\bibnamefont {Lin}}, \bibinfo {author} {\bibfnamefont {C.~H.}\ \bibnamefont
			{Bao}}, \bibinfo {author} {\bibfnamefont {H.~Y.}\ \bibnamefont {Zhang}},
		\bibinfo {author} {\bibfnamefont {S.~N.}\ \bibnamefont {Xu}}, \bibinfo
		{author} {\bibfnamefont {R.~F.}\ \bibnamefont {Feng}}, \bibinfo {author}
		{\bibfnamefont {S.~C.}\ \bibnamefont {Shen}}, \bibinfo {author}
		{\bibfnamefont {K.}~\bibnamefont {Zhao}}, \bibinfo {author} {\bibfnamefont
			{W.}~\bibnamefont {Yao}}, \bibinfo {author} {\bibfnamefont {Y.}~\bibnamefont
			{Wu}}, \bibinfo {author} {\bibfnamefont {S.~H.}\ \bibnamefont {Ji}}, \bibinfo
		{author} {\bibfnamefont {X.}~\bibnamefont {Chen}}, \bibinfo {author}
		{\bibfnamefont {P.~H.}\ \bibnamefont {Tan}}, \bibinfo {author} {\bibfnamefont
			{Q.~K.}\ \bibnamefont {Xue}}, \bibinfo {author} {\bibfnamefont
			{Y.}~\bibnamefont {Xu}}, \bibinfo {author} {\bibfnamefont {W.~H.}\
			\bibnamefont {Duan}}, \bibinfo {author} {\bibfnamefont {P.}~\bibnamefont
			{Yu}}, \ and\ \bibinfo {author} {\bibfnamefont {S.~Y.}\ \bibnamefont
			{Zhou}},\ }\bibfield  {title} {\enquote {\bibinfo {title} {{Tailored Ising
					superconductivity in intercalated bulk NbSe$_2$}},}\ }\href {\doibase
		10.1038/s41567-022-01778-7} {\bibfield  {journal} {\bibinfo  {journal} {Nat.
				Phys.}\ }\textbf {\bibinfo {volume} {18}},\ \bibinfo {pages} {1425} (\bibinfo
		{year} {2022})}\BibitemShut {NoStop}%
	\bibitem [{\citenamefont {Cui}\ \emph {et~al.}(2018)\citenamefont {Cui},
		\citenamefont {Zhang}, \citenamefont {Li}, \citenamefont {Lin}, \citenamefont
		{Zhu}, \citenamefont {Wen}, \citenamefont {Wang}, \citenamefont {Sun},
		\citenamefont {Ma}, \citenamefont {Li}, \citenamefont {Gong}, \citenamefont
		{Xie}, \citenamefont {Gu}, \citenamefont {Li}, \citenamefont {Luo},
		\citenamefont {Yu},\ and\ \citenamefont {Yu}}]{cui_ScienceBulletin_2018}%
	\BibitemOpen
	\bibfield  {author} {\bibinfo {author} {\bibfnamefont {Y.}~\bibnamefont
			{Cui}}, \bibinfo {author} {\bibfnamefont {G.~H.}\ \bibnamefont {Zhang}},
		\bibinfo {author} {\bibfnamefont {H.~B.}\ \bibnamefont {Li}}, \bibinfo
		{author} {\bibfnamefont {H.}~\bibnamefont {Lin}}, \bibinfo {author}
		{\bibfnamefont {X.~Y.}\ \bibnamefont {Zhu}}, \bibinfo {author} {\bibfnamefont
			{H.-H.}\ \bibnamefont {Wen}}, \bibinfo {author} {\bibfnamefont {G.~Q.}\
			\bibnamefont {Wang}}, \bibinfo {author} {\bibfnamefont {J.~Z.}\ \bibnamefont
			{Sun}}, \bibinfo {author} {\bibfnamefont {M.~W.}\ \bibnamefont {Ma}},
		\bibinfo {author} {\bibfnamefont {Y.}~\bibnamefont {Li}}, \bibinfo {author}
		{\bibfnamefont {D.~L.}\ \bibnamefont {Gong}}, \bibinfo {author}
		{\bibfnamefont {T.}~\bibnamefont {Xie}}, \bibinfo {author} {\bibfnamefont
			{Y.~H.}\ \bibnamefont {Gu}}, \bibinfo {author} {\bibfnamefont {S.~L.}\
			\bibnamefont {Li}}, \bibinfo {author} {\bibfnamefont {H.~Q.}\ \bibnamefont
			{Luo}}, \bibinfo {author} {\bibfnamefont {P.}~\bibnamefont {Yu}}, \ and\
		\bibinfo {author} {\bibfnamefont {W.Q.}\ \bibnamefont {Yu}},\ }\bibfield
	{title} {\enquote {\bibinfo {title} {Protonation induced high-$t_c$ phases in
				iron-based superconductors evidenced by $\mathrm{NMR}$ and magnetization
				measurements},}\ }\href {\doibase https://doi.org/10.1016/j.scib.2017.12.009}
	{\bibfield  {journal} {\bibinfo  {journal} {Sci. Bull.}\ }\textbf {\bibinfo
			{volume} {63}},\ \bibinfo {pages} {11} (\bibinfo {year} {2018})}\BibitemShut
	{NoStop}%
	\bibitem [{\citenamefont {Cui}\ \emph {et~al.}(2019)\citenamefont {Cui},
		\citenamefont {Hu}, \citenamefont {Zhang}, \citenamefont {Ma}, \citenamefont
		{Ma}, \citenamefont {Ma}, \citenamefont {Wang}, \citenamefont {Yan},
		\citenamefont {Sun}, \citenamefont {Cheng}, \citenamefont {Jia},
		\citenamefont {Li}, \citenamefont {Wen}, \citenamefont {Lei}, \citenamefont
		{Yu}, \citenamefont {Ji},\ and\ \citenamefont {Yu}}]{Cui_CPL_2019}%
	\BibitemOpen
	\bibfield  {author} {\bibinfo {author} {\bibfnamefont {Y.}~\bibnamefont
			{Cui}}, \bibinfo {author} {\bibfnamefont {Z.}~\bibnamefont {Hu}}, \bibinfo
		{author} {\bibfnamefont {J.~S.}\ \bibnamefont {Zhang}}, \bibinfo {author}
		{\bibfnamefont {W.~L.}\ \bibnamefont {Ma}}, \bibinfo {author} {\bibfnamefont
			{M.~W.}\ \bibnamefont {Ma}}, \bibinfo {author} {\bibfnamefont
			{Z.}~\bibnamefont {Ma}}, \bibinfo {author} {\bibfnamefont {C.}~\bibnamefont
			{Wang}}, \bibinfo {author} {\bibfnamefont {J.~Q.}\ \bibnamefont {Yan}},
		\bibinfo {author} {\bibfnamefont {J.~P.}\ \bibnamefont {Sun}}, \bibinfo
		{author} {\bibfnamefont {J.~G.}\ \bibnamefont {Cheng}}, \bibinfo {author}
		{\bibfnamefont {S.}~\bibnamefont {Jia}}, \bibinfo {author} {\bibfnamefont
			{Y.}~\bibnamefont {Li}}, \bibinfo {author} {\bibfnamefont {J.~S.}\
			\bibnamefont {Wen}}, \bibinfo {author} {\bibfnamefont {H.~C.}\ \bibnamefont
			{Lei}}, \bibinfo {author} {\bibfnamefont {P.}~\bibnamefont {Yu}}, \bibinfo
		{author} {\bibfnamefont {W.}~\bibnamefont {Ji}}, \ and\ \bibinfo {author}
		{\bibfnamefont {W.~Q.}\ \bibnamefont {Yu}},\ }\bibfield  {title} {\enquote
		{\bibinfo {title} {Ionic-liquid-gating induced protonation and
				superconductivity in $\mathrm{FeSe}$,
				$\mathrm{FeSe}_{0.93}\mathrm{S}_{0.07}$, $\mathrm{ZrNCl}$, 1$\mathrm{T}$
				-$\mathrm{TaS}_2$ and $\mathrm{BiSe}_2$},}\ }\href {\doibase
		10.1088/0256-307X/36/7/077401} {\bibfield  {journal} {\bibinfo  {journal}
			{Chin. Phys. Lett}\ }\textbf {\bibinfo {volume} {36}},\ \bibinfo {eid}
		{077401} (\bibinfo {year} {2019})}\BibitemShut {NoStop}%
	\bibitem [{\citenamefont {Mi}\ \emph {et~al.}(2022)\citenamefont {Mi},
		\citenamefont {Zheng}, \citenamefont {Wang}, \citenamefont {Zhou},
		\citenamefont {Yu}, \citenamefont {Xiao}, \citenamefont {Song}, \citenamefont
		{Shen}, \citenamefont {Li}, \citenamefont {Bai}, \citenamefont {Chen},
		\citenamefont {Wang}, \citenamefont {Liu},\ and\ \citenamefont
		{Wang}}]{Mi_AdvFM_2022}%
	\BibitemOpen
	\bibfield  {author} {\bibinfo {author} {\bibfnamefont {M.~J.}\ \bibnamefont
			{Mi}}, \bibinfo {author} {\bibfnamefont {X.~W.}\ \bibnamefont {Zheng}},
		\bibinfo {author} {\bibfnamefont {S.~L.}\ \bibnamefont {Wang}}, \bibinfo
		{author} {\bibfnamefont {Y.}~\bibnamefont {Zhou}}, \bibinfo {author}
		{\bibfnamefont {L.~X.}\ \bibnamefont {Yu}}, \bibinfo {author} {\bibfnamefont
			{H.}~\bibnamefont {Xiao}}, \bibinfo {author} {\bibfnamefont {H.~N.}\
			\bibnamefont {Song}}, \bibinfo {author} {\bibfnamefont {B.}~\bibnamefont
			{Shen}}, \bibinfo {author} {\bibfnamefont {F.~S.}\ \bibnamefont {Li}},
		\bibinfo {author} {\bibfnamefont {L.~H.}\ \bibnamefont {Bai}}, \bibinfo
		{author} {\bibfnamefont {Y.~X.}\ \bibnamefont {Chen}}, \bibinfo {author}
		{\bibfnamefont {S.~P.}\ \bibnamefont {Wang}}, \bibinfo {author}
		{\bibfnamefont {X.~H.}\ \bibnamefont {Liu}}, \ and\ \bibinfo {author}
		{\bibfnamefont {Y.~L.}\ \bibnamefont {Wang}},\ }\bibfield  {title} {\enquote
		{\bibinfo {title} {{Variation between Antiferromagnetism and Ferrimagnetism
					in $\mathrm{NiPS}_3$ by Electron Doping}},}\ }\href {\doibase
		https://doi.org/10.1002/adfm.202112750} {\bibfield  {journal} {\bibinfo
			{journal} {Adv. Funct. Mater.}\ }\textbf {\bibinfo {volume} {32}},\ \bibinfo
		{pages} {2112750} (\bibinfo {year} {2022})}\BibitemShut {NoStop}%
	\bibitem [{\citenamefont {Brec}\ \emph {et~al.}(1985)\citenamefont {Brec},
		\citenamefont {Ouvrard},\ and\ \citenamefont {Rouxel}}]{brec_MRB_1985}%
	\BibitemOpen
	\bibfield  {author} {\bibinfo {author} {\bibfnamefont {R.}~\bibnamefont
			{Brec}}, \bibinfo {author} {\bibfnamefont {G.}~\bibnamefont {Ouvrard}}, \
		and\ \bibinfo {author} {\bibfnamefont {J.}~\bibnamefont {Rouxel}},\
	}\bibfield  {title} {\enquote {\bibinfo {title} {Relationship between
				structure parameters and chemical properties in some $\mathrm{MPS}_3$ layered
				phases},}\ }\href {\doibase https://doi.org/10.1016/0025-5408(85)90118-7}
	{\bibfield  {journal} {\bibinfo  {journal} {Mater. Res. Bull.}\ }\textbf
		{\bibinfo {volume} {20}},\ \bibinfo {pages} {1257} (\bibinfo {year}
		{1985})}\BibitemShut {NoStop}%
	\bibitem [{\citenamefont {Yu}\ \emph {et~al.}(2019)\citenamefont {Yu},
		\citenamefont {Peng}, \citenamefont {Liu}, \citenamefont {Liu}, \citenamefont
		{Liu},\ and\ \citenamefont {Guo}}]{Guo_JMCA_2019}%
	\BibitemOpen
	\bibfield  {author} {\bibinfo {author} {\bibfnamefont {Z.}~\bibnamefont
			{Yu}}, \bibinfo {author} {\bibfnamefont {J.}~\bibnamefont {Peng}}, \bibinfo
		{author} {\bibfnamefont {Y.~H.}\ \bibnamefont {Liu}}, \bibinfo {author}
		{\bibfnamefont {W.~X.}\ \bibnamefont {Liu}}, \bibinfo {author} {\bibfnamefont
			{H.~F.}\ \bibnamefont {Liu}}, \ and\ \bibinfo {author} {\bibfnamefont
			{Y.~Q.}\ \bibnamefont {Guo}},\ }\bibfield  {title} {\enquote {\bibinfo
			{title} {{Amine-assisted exfoliation and electrical conductivity modulation
					toward few-layer $\mathrm{FePS}_3$ nanosheets for efficient hydrogen
					evolution}},}\ }\href {\doibase 10.1039/C9TA03256H} {\bibfield  {journal}
		{\bibinfo  {journal} {J. Mater. Chem. A}\ }\textbf {\bibinfo {volume} {7}},\
		\bibinfo {pages} {13928--13934} (\bibinfo {year} {2019})}\BibitemShut
	{NoStop}%
	\bibitem [{\citenamefont {Wildes}\ \emph {et~al.}(2012)\citenamefont {Wildes},
		\citenamefont {Rule}, \citenamefont {Bewley}, \citenamefont {Enderle},\ and\
		\citenamefont {Hicks}}]{Wildes_JPCM_2012}%
	\BibitemOpen
	\bibfield  {author} {\bibinfo {author} {\bibfnamefont {A.~R.}\ \bibnamefont
			{Wildes}}, \bibinfo {author} {\bibfnamefont {K.~C.}\ \bibnamefont {Rule}},
		\bibinfo {author} {\bibfnamefont {R.~I.}\ \bibnamefont {Bewley}}, \bibinfo
		{author} {\bibfnamefont {M.}~\bibnamefont {Enderle}}, \ and\ \bibinfo
		{author} {\bibfnamefont {T.~J.}\ \bibnamefont {Hicks}},\ }\bibfield  {title}
	{\enquote {\bibinfo {title} {The magnon dynamics and spin exchange parameters
				of $\mathrm{FePS}_3$},}\ }\href {\doibase 10.1088/0953-8984/24/41/416004}
	{\bibfield  {journal} {\bibinfo  {journal} {J. Phys. Condens. Matter}\
		}\textbf {\bibinfo {volume} {24}},\ \bibinfo {pages} {416004} (\bibinfo
		{year} {2012})}\BibitemShut {NoStop}%
	\bibitem [{\citenamefont {Lan\ifmmode~\mbox{\c{c}}\else \c{c}\fi{}on}\ \emph
		{et~al.}(2016)\citenamefont {Lan\ifmmode~\mbox{\c{c}}\else \c{c}\fi{}on},
		\citenamefont {Walker}, \citenamefont {Ressouche}, \citenamefont {Ouladdiaf},
		\citenamefont {Rule}, \citenamefont {McIntyre}, \citenamefont {Hicks},
		\citenamefont {R\o{}nnow},\ and\ \citenamefont {Wildes}}]{Wildes_PRB_2016}%
	\BibitemOpen
	\bibfield  {author} {\bibinfo {author} {\bibfnamefont {D.}~\bibnamefont
			{Lan\ifmmode~\mbox{\c{c}}\else \c{c}\fi{}on}}, \bibinfo {author}
		{\bibfnamefont {H.~C.}\ \bibnamefont {Walker}}, \bibinfo {author}
		{\bibfnamefont {E.}~\bibnamefont {Ressouche}}, \bibinfo {author}
		{\bibfnamefont {B.}~\bibnamefont {Ouladdiaf}}, \bibinfo {author}
		{\bibfnamefont {K.~C.}\ \bibnamefont {Rule}}, \bibinfo {author}
		{\bibfnamefont {G.~J.}\ \bibnamefont {McIntyre}}, \bibinfo {author}
		{\bibfnamefont {T.~J.}\ \bibnamefont {Hicks}}, \bibinfo {author}
		{\bibfnamefont {H.~M.}\ \bibnamefont {R\o{}nnow}}, \ and\ \bibinfo {author}
		{\bibfnamefont {A.~R.}\ \bibnamefont {Wildes}},\ }\bibfield  {title}
	{\enquote {\bibinfo {title} {Magnetic structure and magnon dynamics of the
				quasi-two-dimensional antiferromagnet $\mathrm{FePS}_3$},}\ }\href {\doibase
		10.1103/PhysRevB.94.214407} {\bibfield  {journal} {\bibinfo  {journal} {Phys.
				Rev. B}\ }\textbf {\bibinfo {volume} {94}},\ \bibinfo {pages} {214407}
		(\bibinfo {year} {2016})}\BibitemShut {NoStop}%
	\bibitem [{\citenamefont {Coak}\ \emph {et~al.}(2021)\citenamefont {Coak},
		\citenamefont {Jarvis}, \citenamefont {Hamidov}, \citenamefont {Wildes},
		\citenamefont {Paddison}, \citenamefont {Liu}, \citenamefont {Haines},
		\citenamefont {Dang}, \citenamefont {Kichanov}, \citenamefont {Savenko},
		\citenamefont {Lee}, \citenamefont {Kratochv\'{\i}lov\'a}, \citenamefont
		{Klotz}, \citenamefont {Hansen}, \citenamefont {Kozlenko}, \citenamefont
		{Park},\ and\ \citenamefont {Saxena}}]{Coak_PRX_2021}%
	\BibitemOpen
	\bibfield  {author} {\bibinfo {author} {\bibfnamefont {M.~J.}\ \bibnamefont
			{Coak}}, \bibinfo {author} {\bibfnamefont {D.~M.}\ \bibnamefont {Jarvis}},
		\bibinfo {author} {\bibfnamefont {H.}~\bibnamefont {Hamidov}}, \bibinfo
		{author} {\bibfnamefont {A.~R.}\ \bibnamefont {Wildes}}, \bibinfo {author}
		{\bibfnamefont {J.~A.~M.}\ \bibnamefont {Paddison}}, \bibinfo {author}
		{\bibfnamefont {C.}~\bibnamefont {Liu}}, \bibinfo {author} {\bibfnamefont
			{C.~R.~S.}\ \bibnamefont {Haines}}, \bibinfo {author} {\bibfnamefont {N.~T.}\
			\bibnamefont {Dang}}, \bibinfo {author} {\bibfnamefont {S.~E.}\ \bibnamefont
			{Kichanov}}, \bibinfo {author} {\bibfnamefont {B.~N.}\ \bibnamefont
			{Savenko}}, \bibinfo {author} {\bibfnamefont {S.}~\bibnamefont {Lee}},
		\bibinfo {author} {\bibfnamefont {M.}~\bibnamefont {Kratochv\'{\i}lov\'a}},
		\bibinfo {author} {\bibfnamefont {S.}~\bibnamefont {Klotz}}, \bibinfo
		{author} {\bibfnamefont {T.~C.}\ \bibnamefont {Hansen}}, \bibinfo {author}
		{\bibfnamefont {D.~P.}\ \bibnamefont {Kozlenko}}, \bibinfo {author}
		{\bibfnamefont {J.~G}\ \bibnamefont {Park}}, \ and\ \bibinfo {author}
		{\bibfnamefont {S.~S.}\ \bibnamefont {Saxena}},\ }\bibfield  {title}
	{\enquote {\bibinfo {title} {{Emergent Magnetic Phases in Pressure-Tuned van
					der Waals Antiferromagnet $\mathrm{FePS}_{3}$}},}\ }\href {\doibase
		10.1103/PhysRevX.11.011024} {\bibfield  {journal} {\bibinfo  {journal} {Phys.
				Rev. X}\ }\textbf {\bibinfo {volume} {11}},\ \bibinfo {pages} {011024}
		(\bibinfo {year} {2021})}\BibitemShut {NoStop}%
	\bibitem [{\citenamefont {Paul}\ \emph {et~al.}(2023)\citenamefont {Paul},
		\citenamefont {Negi}, \citenamefont {Talukdar}, \citenamefont {Karak},
		\citenamefont {Badola}, \citenamefont {Poojitha}, \citenamefont {Mandal},
		\citenamefont {Marik}, \citenamefont {Singh}, \citenamefont {Pistawala},
		\citenamefont {Harnagea}, \citenamefont {Thomas}, \citenamefont {Soni},
		\citenamefont {Bhattacharjee},\ and\ \citenamefont
		{Saha}}]{paul_2023_tuning}%
	\BibitemOpen
	\bibfield  {author} {\bibinfo {author} {\bibfnamefont {S.}~\bibnamefont
			{Paul}}, \bibinfo {author} {\bibfnamefont {D.}~\bibnamefont {Negi}}, \bibinfo
		{author} {\bibfnamefont {S.}~\bibnamefont {Talukdar}}, \bibinfo {author}
		{\bibfnamefont {S.}~\bibnamefont {Karak}}, \bibinfo {author} {\bibfnamefont
			{S.}~\bibnamefont {Badola}}, \bibinfo {author} {\bibfnamefont
			{B.}~\bibnamefont {Poojitha}}, \bibinfo {author} {\bibfnamefont
			{M.}~\bibnamefont {Mandal}}, \bibinfo {author} {\bibfnamefont
			{S.}~\bibnamefont {Marik}}, \bibinfo {author} {\bibfnamefont {R.~P.}\
			\bibnamefont {Singh}}, \bibinfo {author} {\bibfnamefont {N.}~\bibnamefont
			{Pistawala}}, \bibinfo {author} {\bibfnamefont {L.}~\bibnamefont {Harnagea}},
		\bibinfo {author} {\bibfnamefont {A.}~\bibnamefont {Thomas}}, \bibinfo
		{author} {\bibfnamefont {A.}~\bibnamefont {Soni}}, \bibinfo {author}
		{\bibfnamefont {S.}~\bibnamefont {Bhattacharjee}}, \ and\ \bibinfo {author}
		{\bibfnamefont {S.}~\bibnamefont {Saha}},\ }\bibfield  {title} {\enquote
		{\bibinfo {title} {{Tuning the magnetic properties in $\mathrm{MPS}_3$
					($\mathrm{M}$ = $\mathrm{Mn}$, $\mathrm{Fe}$ and $\mathrm{Ni}$) by
					proximity-induced Dzyaloshinskii-Moriya interactions}},}\ }\href@noop {} {\
		(\bibinfo {year} {2023})},\ \Eprint {http://arxiv.org/abs/2307.13400}
	{arXiv:2307.13400} \BibitemShut {NoStop}%
	\bibitem [{\citenamefont {Wildes}\ \emph {et~al.}(2015)\citenamefont {Wildes},
		\citenamefont {Simonet}, \citenamefont {Ressouche}, \citenamefont {McIntyre},
		\citenamefont {Avdeev}, \citenamefont {Suard}, \citenamefont {Kimber},
		\citenamefont {Lan\ifmmode~\mbox{\c{c}}\else \c{c}\fi{}on}, \citenamefont
		{Pepe}, \citenamefont {Moubaraki},\ and\ \citenamefont
		{Hicks}}]{Wildes_PRB_2015}%
	\BibitemOpen
	\bibfield  {author} {\bibinfo {author} {\bibfnamefont {A.~R.}\ \bibnamefont
			{Wildes}}, \bibinfo {author} {\bibfnamefont {V.}~\bibnamefont {Simonet}},
		\bibinfo {author} {\bibfnamefont {E.}~\bibnamefont {Ressouche}}, \bibinfo
		{author} {\bibfnamefont {G.~J.}\ \bibnamefont {McIntyre}}, \bibinfo {author}
		{\bibfnamefont {M.}~\bibnamefont {Avdeev}}, \bibinfo {author} {\bibfnamefont
			{E.}~\bibnamefont {Suard}}, \bibinfo {author} {\bibfnamefont {S.~A.~J.}\
			\bibnamefont {Kimber}}, \bibinfo {author} {\bibfnamefont {D.}~\bibnamefont
			{Lan\ifmmode~\mbox{\c{c}}\else \c{c}\fi{}on}}, \bibinfo {author}
		{\bibfnamefont {G.}~\bibnamefont {Pepe}}, \bibinfo {author} {\bibfnamefont
			{B.}~\bibnamefont {Moubaraki}}, \ and\ \bibinfo {author} {\bibfnamefont
			{T.~J.}\ \bibnamefont {Hicks}},\ }\bibfield  {title} {\enquote {\bibinfo
			{title} {Magnetic structure of the quasi-two-dimensional antiferromagnet
				$\mathrm{NiPS}_3$},}\ }\href {\doibase 10.1103/PhysRevB.92.224408} {\bibfield
		{journal} {\bibinfo  {journal} {Phys. Rev. B}\ }\textbf {\bibinfo {volume}
			{92}},\ \bibinfo {pages} {224408} (\bibinfo {year} {2015})}\BibitemShut
	{NoStop}%
	\bibitem [{\citenamefont {Kim}\ and\ \citenamefont {Park}(2021)}]{Kim_NL_2021}%
	\BibitemOpen
	\bibfield  {author} {\bibinfo {author} {\bibfnamefont {T.~Y.}\ \bibnamefont
			{Kim}}\ and\ \bibinfo {author} {\bibfnamefont {C.~H.}\ \bibnamefont {Park}},\
	}\bibfield  {title} {\enquote {\bibinfo {title} {Magnetic anisotropy and
				magnetic ordering of transition-metal phosphorus trisulfides},}\ }\href
	{\doibase 10.1021/acs.nanolett.1c03992} {\bibfield  {journal} {\bibinfo
			{journal} {Nano Lett.}\ }\textbf {\bibinfo {volume} {21}},\ \bibinfo {pages}
		{10114} (\bibinfo {year} {2021})}\BibitemShut {NoStop}%
	\bibitem [{\citenamefont {Wildes}\ \emph {et~al.}(2022)\citenamefont {Wildes},
		\citenamefont {Stewart}, \citenamefont {Le}, \citenamefont {Ewings},
		\citenamefont {Rule}, \citenamefont {Deng},\ and\ \citenamefont
		{Anand}}]{Wildes_PRB_2022}%
	\BibitemOpen
	\bibfield  {author} {\bibinfo {author} {\bibfnamefont {A.~R.}\ \bibnamefont
			{Wildes}}, \bibinfo {author} {\bibfnamefont {J.~R.}\ \bibnamefont {Stewart}},
		\bibinfo {author} {\bibfnamefont {M.~D.}\ \bibnamefont {Le}}, \bibinfo
		{author} {\bibfnamefont {R.~A.}\ \bibnamefont {Ewings}}, \bibinfo {author}
		{\bibfnamefont {K.~C.}\ \bibnamefont {Rule}}, \bibinfo {author}
		{\bibfnamefont {G.}~\bibnamefont {Deng}}, \ and\ \bibinfo {author}
		{\bibfnamefont {K.}~\bibnamefont {Anand}},\ }\bibfield  {title} {\enquote
		{\bibinfo {title} {Magnetic dynamics of $\mathrm{NiPS}_{3}$},}\ }\href
	{\doibase 10.1103/PhysRevB.106.174422} {\bibfield  {journal} {\bibinfo
			{journal} {Phys. Rev. B}\ }\textbf {\bibinfo {volume} {106}},\ \bibinfo
		{pages} {174422} (\bibinfo {year} {2022})}\BibitemShut {NoStop}%
	\bibitem [{\citenamefont {Amirabbasi}\ and\ \citenamefont
		{Kratzer}(2023)}]{Peter_PRB_2023}%
	\BibitemOpen
	\bibfield  {author} {\bibinfo {author} {\bibfnamefont {M.}~\bibnamefont
			{Amirabbasi}}\ and\ \bibinfo {author} {\bibfnamefont {P.}~\bibnamefont
			{Kratzer}},\ }\bibfield  {title} {\enquote {\bibinfo {title} {Orbital and
				magnetic ordering in single-layer $\mathrm{FePS}_3$: A
				$\mathrm{DFT}+\mathrm{U}$ study},}\ }\href {\doibase
		10.1103/PhysRevB.107.024401} {\bibfield  {journal} {\bibinfo  {journal}
			{Phys. Rev. B}\ }\textbf {\bibinfo {volume} {107}},\ \bibinfo {pages}
		{024401} (\bibinfo {year} {2023})}\BibitemShut {NoStop}%
	\bibitem [{\citenamefont {Okuda}\ \emph {et~al.}(1983)\citenamefont {Okuda},
		\citenamefont {Kurosawa},\ and\ \citenamefont {Saito}}]{Okuda_1983}%
	\BibitemOpen
	\bibfield  {author} {\bibinfo {author} {\bibfnamefont {K.}~\bibnamefont
			{Okuda}}, \bibinfo {author} {\bibfnamefont {K.}~\bibnamefont {Kurosawa}}, \
		and\ \bibinfo {author} {\bibfnamefont {S.}~\bibnamefont {Saito}},\ }\bibfield
	{title} {\enquote {\bibinfo {title} {High field magnetization process in
				$\mathrm{FePS}_3$},}\ }in\ \href {\doibase
		https://doi.org/10.1016/B978-0-444-86566-3.50011-1} {\emph {\bibinfo
			{booktitle} {High Field Magnetism}}},\ \bibinfo {editor} {edited by\ \bibinfo
		{editor} {\bibfnamefont {M.}~\bibnamefont {Date}}}\ (\bibinfo  {publisher}
	{Elsevier},\ \bibinfo {address} {Amsterdam},\ \bibinfo {year} {1983})\ pp.\
	\bibinfo {pages} {55--58}\BibitemShut {NoStop}%
	\bibitem [{\citenamefont {Jernberg}\ \emph {et~al.}(1984)\citenamefont
		{Jernberg}, \citenamefont {Bjarman},\ and\ \citenamefont
		{W\"appling}}]{Jernberg_JM_1984}%
	\BibitemOpen
	\bibfield  {author} {\bibinfo {author} {\bibfnamefont {P.}~\bibnamefont
			{Jernberg}}, \bibinfo {author} {\bibfnamefont {S.}~\bibnamefont {Bjarman}}, \
		and\ \bibinfo {author} {\bibfnamefont {R.}~\bibnamefont {W\"appling}},\
	}\bibfield  {title} {\enquote {\bibinfo {title} {$\mathrm{FePS}_3$: A
				first-order phase transition in 2$\mathrm{D}$ $\mathrm{I}$sing
				antiferromagnet},}\ }\href {\doibase
		https://doi.org/10.1016/0304-8853(84)90355-X} {\bibfield  {journal} {\bibinfo
			{journal} {J. Magn. Magn. Mater.}\ }\textbf {\bibinfo {volume} {46}},\
		\bibinfo {pages} {178} (\bibinfo {year} {1984})}\BibitemShut {NoStop}%
	\bibitem [{\citenamefont {Ferloni}\ and\ \citenamefont
		{Scagliotti}(1989)}]{Ferloni_ThermochimicaActa_1989}%
	\BibitemOpen
	\bibfield  {author} {\bibinfo {author} {\bibfnamefont {P.}~\bibnamefont
			{Ferloni}}\ and\ \bibinfo {author} {\bibfnamefont {M.}~\bibnamefont
			{Scagliotti}},\ }\bibfield  {title} {\enquote {\bibinfo {title} {Magnetic
				phase transitions in iron and nickel phosphorus trichalcogenides},}\ }\href
	{\doibase https://doi.org/10.1016/0040-6031(89)87022-4} {\bibfield  {journal}
		{\bibinfo  {journal} {Thermochim. Acta}\ }\textbf {\bibinfo {volume} {139}},\
		\bibinfo {pages} {197} (\bibinfo {year} {1989})}\BibitemShut {NoStop}%
	\bibitem [{\citenamefont {Ziolo}\ \emph {et~al.}(1988)\citenamefont {Ziolo},
		\citenamefont {Torre}, \citenamefont {Rigamonti},\ and\ \citenamefont
		{Borsa}}]{Ziolo_JAP_1988}%
	\BibitemOpen
	\bibfield  {author} {\bibinfo {author} {\bibfnamefont {J.}~\bibnamefont
			{Ziolo}}, \bibinfo {author} {\bibfnamefont {S.}~\bibnamefont {Torre}},
		\bibinfo {author} {\bibfnamefont {A.}~\bibnamefont {Rigamonti}}, \ and\
		\bibinfo {author} {\bibfnamefont {F.}~\bibnamefont {Borsa}},\ }\bibfield
	{title} {\enquote {\bibinfo {title} {{$^{31}$P NMR relaxation study of spin
					dynamics in layered transition-metal compounds $\mathrm{MPX}_3$}},}\ }\href
	{\doibase 10.1063/1.341169} {\bibfield  {journal} {\bibinfo  {journal} {J.
				Appl. Phys.}\ }\textbf {\bibinfo {volume} {63}},\ \bibinfo {pages} {3095}
		(\bibinfo {year} {1988})}\BibitemShut {NoStop}%
	\bibitem [{\citenamefont {Rosenblum}\ \emph {et~al.}(1994)\citenamefont
		{Rosenblum}, \citenamefont {Francis},\ and\ \citenamefont
		{Merlin}}]{Rosen_PRB_1994}%
	\BibitemOpen
	\bibfield  {author} {\bibinfo {author} {\bibfnamefont {S.}~\bibnamefont
			{Rosenblum}}, \bibinfo {author} {\bibfnamefont {A.~H.}\ \bibnamefont
			{Francis}}, \ and\ \bibinfo {author} {\bibfnamefont {R.}~\bibnamefont
			{Merlin}},\ }\bibfield  {title} {\enquote {\bibinfo {title} {Two-magnon light
				scattering in the layered antiferromagnet $\mathrm{NiPS}_3$:
				$\mathrm{Spin}$-1/2-like anomalies in a spin-1 system},}\ }\href {\doibase
		10.1103/PhysRevB.49.4352} {\bibfield  {journal} {\bibinfo  {journal} {Phys.
				Rev. B}\ }\textbf {\bibinfo {volume} {49}},\ \bibinfo {pages} {4352}
		(\bibinfo {year} {1994})}\BibitemShut {NoStop}%
	\bibitem [{\citenamefont {Kim}\ \emph {et~al.}(2019)\citenamefont {Kim},
		\citenamefont {Lim}, \citenamefont {Lee}, \citenamefont {Lee}, \citenamefont
		{Kim}, \citenamefont {Park}, \citenamefont {Jeon}, \citenamefont {Park},
		\citenamefont {Park},\ and\ \citenamefont {Cheong}}]{Kim_NC_2019}%
	\BibitemOpen
	\bibfield  {author} {\bibinfo {author} {\bibfnamefont {K.}~\bibnamefont
			{Kim}}, \bibinfo {author} {\bibfnamefont {S.~Y.}\ \bibnamefont {Lim}},
		\bibinfo {author} {\bibfnamefont {J.-U.}\ \bibnamefont {Lee}}, \bibinfo
		{author} {\bibfnamefont {S.}~\bibnamefont {Lee}}, \bibinfo {author}
		{\bibfnamefont {T.~Y.}\ \bibnamefont {Kim}}, \bibinfo {author} {\bibfnamefont
			{K.}~\bibnamefont {Park}}, \bibinfo {author} {\bibfnamefont {G.~S.}\
			\bibnamefont {Jeon}}, \bibinfo {author} {\bibfnamefont {C.-H.}\ \bibnamefont
			{Park}}, \bibinfo {author} {\bibfnamefont {J.-G.}\ \bibnamefont {Park}}, \
		and\ \bibinfo {author} {\bibfnamefont {H.}~\bibnamefont {Cheong}},\
	}\bibfield  {title} {\enquote {\bibinfo {title} {Suppression of magnetic
				ordering in $\mathrm{XXZ}$-type antiferromagnetic monolayer
				$\mathrm{NiPS}_3$},}\ }\href {\doibase doi.org/10.1038/s41467-018-08284-6}
	{\bibfield  {journal} {\bibinfo  {journal} {Nat. Commun.}\ }\textbf {\bibinfo
			{volume} {10}},\ \bibinfo {pages} {345} (\bibinfo {year} {2019})}\BibitemShut
	{NoStop}%
	\bibitem [{\citenamefont {Jana}\ \emph {et~al.}(2023)\citenamefont {Jana},
		\citenamefont {Kapuscinski}, \citenamefont {Mohelsky}, \citenamefont
		{Vaclavkova}, \citenamefont {Breslavetz}, \citenamefont {Orlita},
		\citenamefont {Faugeras},\ and\ \citenamefont {Potemski}}]{Jana_PRB_2023}%
	\BibitemOpen
	\bibfield  {author} {\bibinfo {author} {\bibfnamefont {D.}~\bibnamefont
			{Jana}}, \bibinfo {author} {\bibfnamefont {P.}~\bibnamefont {Kapuscinski}},
		\bibinfo {author} {\bibfnamefont {I.}~\bibnamefont {Mohelsky}}, \bibinfo
		{author} {\bibfnamefont {D.}~\bibnamefont {Vaclavkova}}, \bibinfo {author}
		{\bibfnamefont {I.}~\bibnamefont {Breslavetz}}, \bibinfo {author}
		{\bibfnamefont {M.}~\bibnamefont {Orlita}}, \bibinfo {author} {\bibfnamefont
			{C.}~\bibnamefont {Faugeras}}, \ and\ \bibinfo {author} {\bibfnamefont
			{M.}~\bibnamefont {Potemski}},\ }\bibfield  {title} {\enquote {\bibinfo
			{title} {{Magnon gap excitations and spin-entangled optical transition in the
					van der Waals antiferromagnet $\mathrm{NiPS}_3$}},}\ }\href {\doibase
		10.1103/PhysRevB.108.115149} {\bibfield  {journal} {\bibinfo  {journal}
			{Phys. Rev. B}\ }\textbf {\bibinfo {volume} {108}},\ \bibinfo {pages}
		{115149} (\bibinfo {year} {2023})}\BibitemShut {NoStop}%
	\bibitem [{\citenamefont {Yasuda}\ \emph {et~al.}(2005)\citenamefont {Yasuda},
		\citenamefont {Todo}, \citenamefont {Hukushima}, \citenamefont {Alet},
		\citenamefont {Keller}, \citenamefont {Troyer},\ and\ \citenamefont
		{Takayama}}]{Yasuda_PRL_2018}%
	\BibitemOpen
	\bibfield  {author} {\bibinfo {author} {\bibfnamefont {C.}~\bibnamefont
			{Yasuda}}, \bibinfo {author} {\bibfnamefont {S.}~\bibnamefont {Todo}},
		\bibinfo {author} {\bibfnamefont {K.}~\bibnamefont {Hukushima}}, \bibinfo
		{author} {\bibfnamefont {F.}~\bibnamefont {Alet}}, \bibinfo {author}
		{\bibfnamefont {M.}~\bibnamefont {Keller}}, \bibinfo {author} {\bibfnamefont
			{M.}~\bibnamefont {Troyer}}, \ and\ \bibinfo {author} {\bibfnamefont
			{H.}~\bibnamefont {Takayama}},\ }\bibfield  {title} {\enquote {\bibinfo
			{title} {{N\'eel Temperature of Quasi-Low-Dimensional Heisenberg
					Antiferromagnets}},}\ }\href {\doibase 10.1103/PhysRevLett.94.217201}
	{\bibfield  {journal} {\bibinfo  {journal} {Phys. Rev. Lett.}\ }\textbf
		{\bibinfo {volume} {94}},\ \bibinfo {pages} {217201} (\bibinfo {year}
		{2005})}\BibitemShut {NoStop}%
	\bibitem [{\citenamefont {Lee}\ \emph {et~al.}(2023)\citenamefont {Lee},
		\citenamefont {Son}, \citenamefont {Kim}, \citenamefont {Kang}, \citenamefont
		{Shen}, \citenamefont {Kenzelmann}, \citenamefont {Delley}, \citenamefont
		{Savchenko}, \citenamefont {Parchenko}, \citenamefont {Na}, \citenamefont
		{Choi}, \citenamefont {Kim}, \citenamefont {Cheong}, \citenamefont {Derlet},
		\citenamefont {Kleibert},\ and\ \citenamefont {Park}}]{Lee_AEM_2023}%
	\BibitemOpen
	\bibfield  {author} {\bibinfo {author} {\bibfnamefont {Y.}~\bibnamefont
			{Lee}}, \bibinfo {author} {\bibfnamefont {S.}~\bibnamefont {Son}}, \bibinfo
		{author} {\bibfnamefont {C.}~\bibnamefont {Kim}}, \bibinfo {author}
		{\bibfnamefont {S.}~\bibnamefont {Kang}}, \bibinfo {author} {\bibfnamefont
			{J.}~\bibnamefont {Shen}}, \bibinfo {author} {\bibfnamefont {M.}~\bibnamefont
			{Kenzelmann}}, \bibinfo {author} {\bibfnamefont {B.}~\bibnamefont {Delley}},
		\bibinfo {author} {\bibfnamefont {T.}~\bibnamefont {Savchenko}}, \bibinfo
		{author} {\bibfnamefont {S.}~\bibnamefont {Parchenko}}, \bibinfo {author}
		{\bibfnamefont {W.}~\bibnamefont {Na}}, \bibinfo {author} {\bibfnamefont
			{K.-Y.}\ \bibnamefont {Choi}}, \bibinfo {author} {\bibfnamefont
			{W.}~\bibnamefont {Kim}}, \bibinfo {author} {\bibfnamefont {H.}~\bibnamefont
			{Cheong}}, \bibinfo {author} {\bibfnamefont {P.~M.}\ \bibnamefont {Derlet}},
		\bibinfo {author} {\bibfnamefont {A.}~\bibnamefont {Kleibert}}, \ and\
		\bibinfo {author} {\bibfnamefont {J.-G.}\ \bibnamefont {Park}},\ }\bibfield
	{title} {\enquote {\bibinfo {title} {{Giant Magnetic Anisotropy in the
					Atomically Thin van der Waals Antiferromagnet FePS$_3$}},}\ }\href {\doibase
		10.1002/aelm.202200650} {\bibfield  {journal} {\bibinfo  {journal} {Adv.
				Electron. Mater.}\ }\textbf {\bibinfo {volume} {9}},\ \bibinfo {pages}
		{2200650} (\bibinfo {year} {2023})}\BibitemShut {NoStop}%
	\bibitem [{\citenamefont {Mertens}\ \emph {et~al.}(2023)\citenamefont
		{Mertens}, \citenamefont {M\"onkeb\"uscher}, \citenamefont {Parlak},
		\citenamefont {Boix-Constant}, \citenamefont {Ma$\Tilde{\rm n}$as-Valero},
		\citenamefont {Matzer}, \citenamefont {Adhikari}, \citenamefont {Bonanni},
		\citenamefont {Coronado}, \citenamefont {Kalashnikova}, \citenamefont
		{Bossini},\ and\ \citenamefont {Cinchetti}}]{Mirko_AM_2023}%
	\BibitemOpen
	\bibfield  {author} {\bibinfo {author} {\bibfnamefont {F.}~\bibnamefont
			{Mertens}}, \bibinfo {author} {\bibfnamefont {D.}~\bibnamefont
			{M\"onkeb\"uscher}}, \bibinfo {author} {\bibfnamefont {U.}~\bibnamefont
			{Parlak}}, \bibinfo {author} {\bibfnamefont {C.}~\bibnamefont
			{Boix-Constant}}, \bibinfo {author} {\bibfnamefont {S.}~\bibnamefont
			{Ma$\Tilde{\rm n}$as-Valero}}, \bibinfo {author} {\bibfnamefont
			{M.}~\bibnamefont {Matzer}}, \bibinfo {author} {\bibfnamefont
			{R.}~\bibnamefont {Adhikari}}, \bibinfo {author} {\bibfnamefont
			{A.}~\bibnamefont {Bonanni}}, \bibinfo {author} {\bibfnamefont
			{E.}~\bibnamefont {Coronado}}, \bibinfo {author} {\bibfnamefont {A.~M.}\
			\bibnamefont {Kalashnikova}}, \bibinfo {author} {\bibfnamefont
			{D.}~\bibnamefont {Bossini}}, \ and\ \bibinfo {author} {\bibfnamefont
			{M.}~\bibnamefont {Cinchetti}},\ }\bibfield  {title} {\enquote {\bibinfo
			{title} {{Ultrafast Coherent THz Lattice Dynamics Coupled to Spins in the van
					der Waals Antiferromagnet $\mathrm{FePS}_3$}},}\ }\href {\doibase
		https://doi.org/10.1002/adma.202208355} {\bibfield  {journal} {\bibinfo
			{journal} {Adv. Mater.}\ }\textbf {\bibinfo {volume} {35}},\ \bibinfo {pages}
		{2208355} (\bibinfo {year} {2023})}\BibitemShut {NoStop}%
	\bibitem [{\citenamefont {Mellado}(2023)}]{Paula_APL_2023}%
	\BibitemOpen
	\bibfield  {author} {\bibinfo {author} {\bibfnamefont {P.}~\bibnamefont
			{Mellado}},\ }\bibfield  {title} {\enquote {\bibinfo {title} {{Spin model for
					the honeycomb NiPS$_3$}},}\ }\href {\doibase 10.1063/5.0176703} {\bibfield
		{journal} {\bibinfo  {journal} {Appl. Phys. Lett.}\ }\textbf {\bibinfo
			{volume} {123}},\ \bibinfo {pages} {242403} (\bibinfo {year}
		{2023})}\BibitemShut {NoStop}%
\end{thebibliography}

%

\end{document}